# Optical Design and Active Optics Methods in Astronomy

Gerard R. LEMAITRE

*Laboratoire d'Astrophysique de Marseille, LOOM, Aix Marseille Université and CNRS,*
*38 rue Fréderic Joliot-Curie, F-13388 Marseille CX 13, France*

Optical designs for astronomy involve implementation of active optics and adaptive optics from X-ray to the infrared. Developments and results of active optics methods for telescopes, spectrographs and coronagraph planet finders are presented. The high accuracy and remarkable smoothness of surfaces generated by active optics methods also allow elaborating new optical design types with high aspheric and/or non-axisymmetric surfaces. Depending on the goal and performance requested for a deformable optical surface analytical investigations are carried out with one of the various facets of elasticity theory: small deformation thin plate theory, large deformation thin plate theory, shallow spherical shell theory, weakly conical shell theory. The resulting thickness distribution and associated bending force boundaries can be refined further with finite element analysis.

## 1. Introduction

This review on optical design and active optics methods presents various concepts of deformable telescope optics which have been elaborated and developed at our optical laboratory—LOOM—, for more than 40 years, and at some other astronomical institutes around the world. Elasticity analyzes and optical designs allow optimizing substrate geometry with appropriate boundary conditions for applying bending forces. For materials behaving a linear stress-strain relationship, as glass and some metal alloys, these methods provide accurate optical deformation modes which satisfy the diffraction-limited criteria.

## 2. Active Optics Procedures: Elasticity

Active optics methods application fields are mainly:

- stress figuring processes of optical surfaces with large asphetization capability,
- in situ shaping processes of telescope mirrors with large asphetization capability,
- in situ reshaping and alignment of large telescope optics with close-loop wavefront sensors,
- variable curvature—or zoom—mirrors for field cophasing of telescope arrays and two-arm interferometers,
- aberration corrected diffraction grating made by replication from actively aspherized submasters,
- universal deformable compensators for photosynthesis holographic recording of corrected gratings.

Applications of elasticity theory to the mirror case led us to consider various substrate classes: constant-thickness, quasi-constant-thickness, variable-thickness and hybrid-thickness classes. The elasticity theory of thin plates provides accurate results when the curvature of the middle surface of the substrate is low whereas the theory of shallow shells is required for more curved substrates, say, for mirrors faster than f/3. The large deformation theory allows accurate designs for variable curvature mirrors with large zoom ranges. Results from the new theory of weakly conical shells are presented for X-ray telescope using tubular mirrors.

A book on *Active Optics Methods* including analytic developments of the elasticity theory have been recently published by the author.[1)]

## 3. Advantages and Origins of Active Optics

To avoid *surface discontinuities* or *ripple errors* caused by local retouches with small figuring laps, B. Schmidt, inventor of the wide-field telescope, suggested, around 1932, that an aspheric surface should be obtained by elastic relaxation after spherical—or plane—figuring with a *full aperture lap*. Schmidt's idea of obtaining smooth aspheric

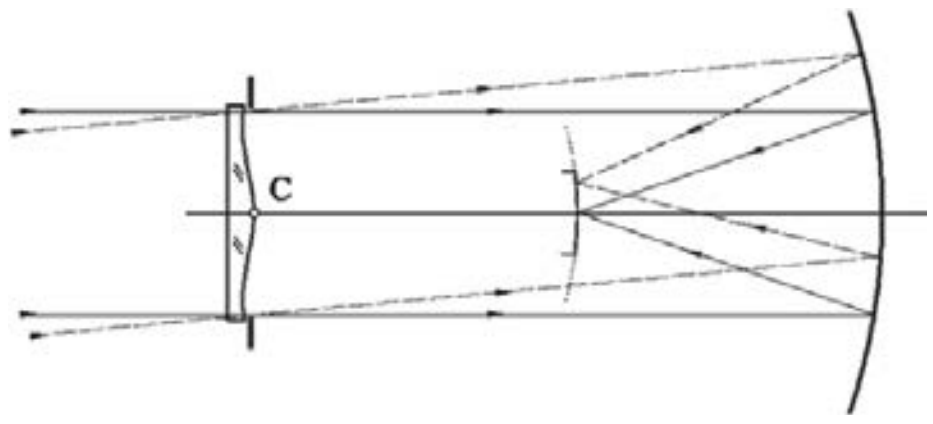

Fig. 1. Schmidt wide-field telescope design.

surfaces—i.e., free from high spatial frequency errors—remained of great potential interest and was also advocated by H. Chretien. The origin of active optics goes back to Everhart[2)] in 1965 who aspherized a refractive corrector plate for a Schmidt telescope (Figs. 1 and 2).

## 4. Stress Aspherization of Refractive Corrector Plates

### *4.1 Single-zone loading method*

Early developments by Everhart,[2)] now called the single zone loading methods (Fig. 3), use all full-size and spherical laps.

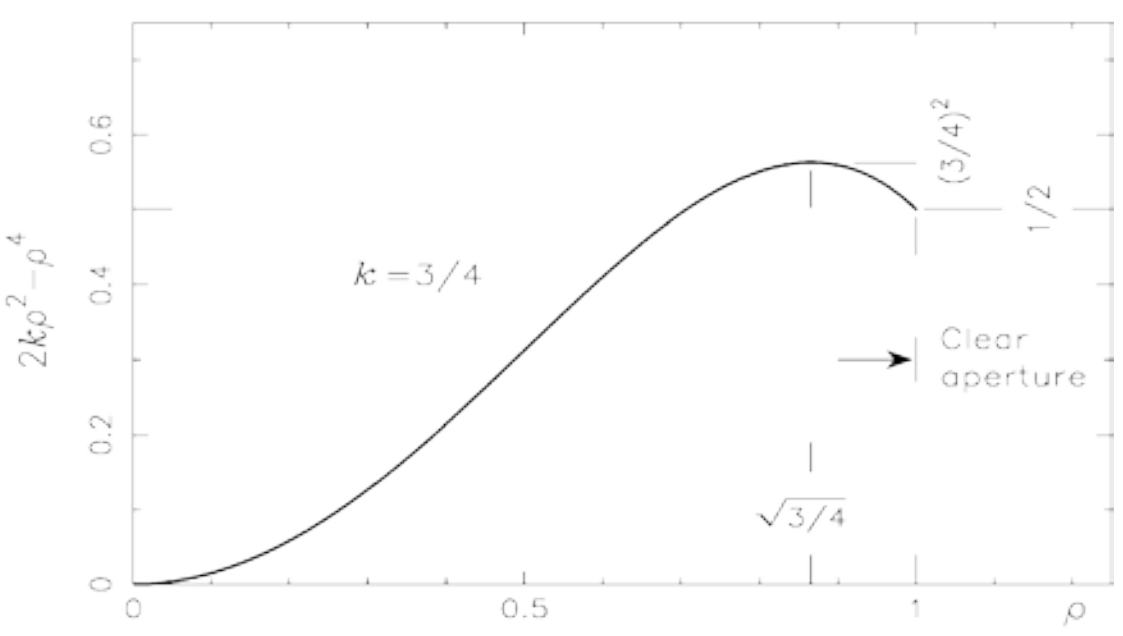

Fig. 2. Kerber profile[1)] of Schmidt corrector plate. Balance of slopes minimizes spherochromatism.

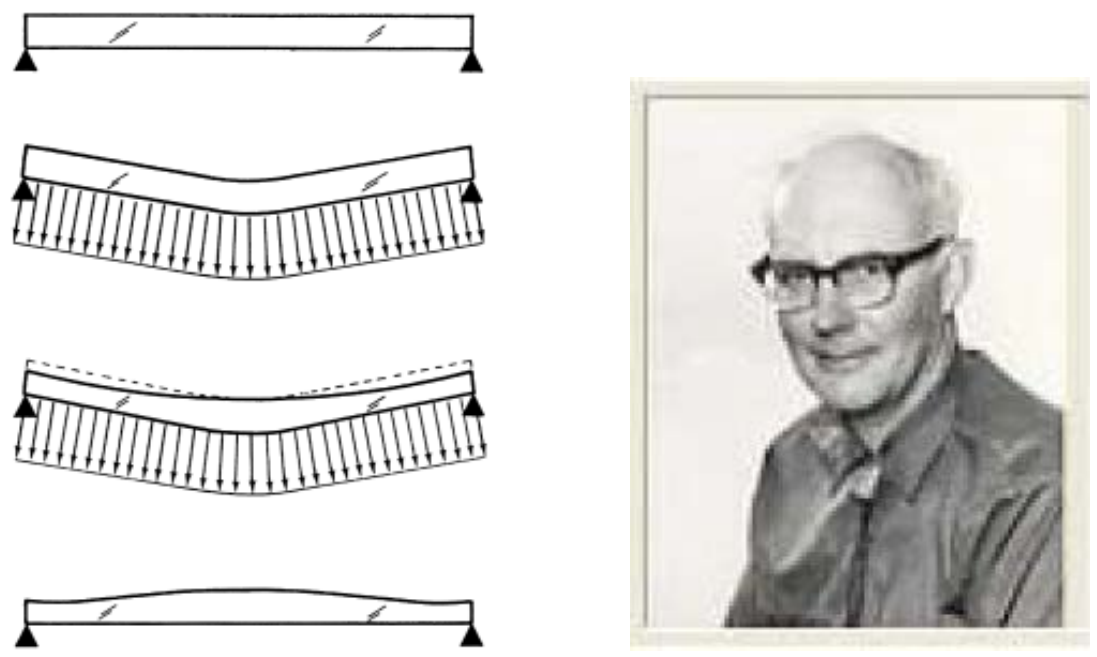

Fig. 3. Plate asphierization. Single-zone method. Portrait of E. Everhart (University of Denver, CO, Penrose Archives).

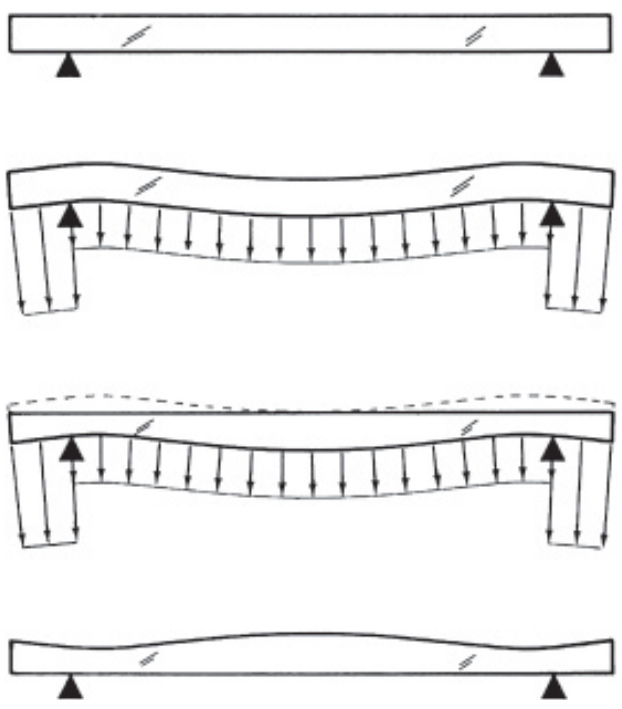

Fig. 4. Plate asphierization. Double-zone method.

By partial vacuum under a plate supported near its edge, and solving Poisson's equation of the flexure $\Delta^2\Delta^2 z(\rho)$ + constant = 0, he aspherized a 29-cm plate for a Schmidt telescope.

*4.2 Double-zone loading method*

Because of technical difficulties in the above method, caused by the lack in control of the lap radius of curvature, and then the control of the null-power zone of the plate at √3 /2 = 0.866, one set up in 1969 the double-zone loading method by Lemaitre[3)] with full-size plane figuring laps (Fig. 4). This method, proposed and developed at LOOM, provided about a hundred corrector plates from f/3.5 up to f/1.1 for telescopes, spectrographs and embarked cameras up to 62-cm clear aperture[3)] (Figs. 5 and 6).

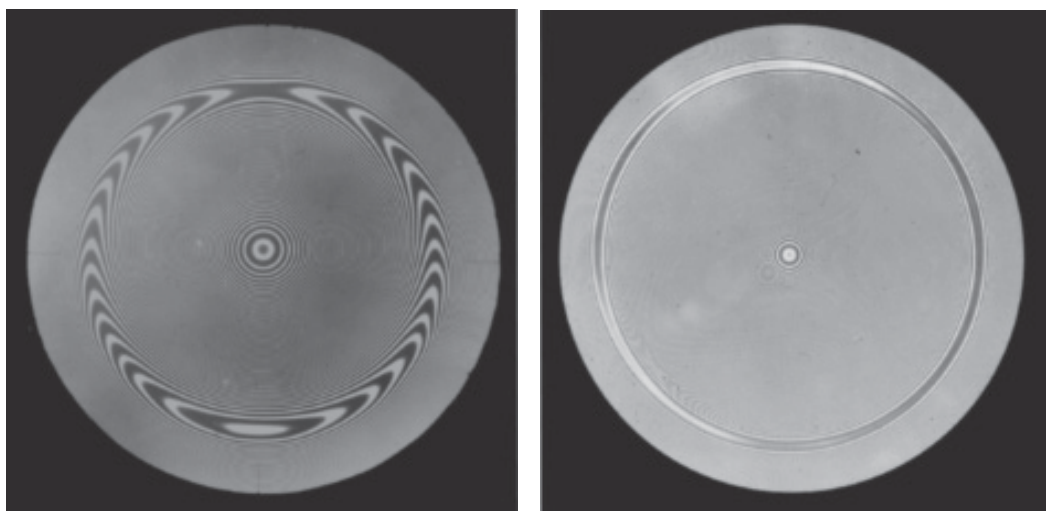

Fig. 5. He–Ne equal-thickness fringes of plates aspherized by *double-zone method*.

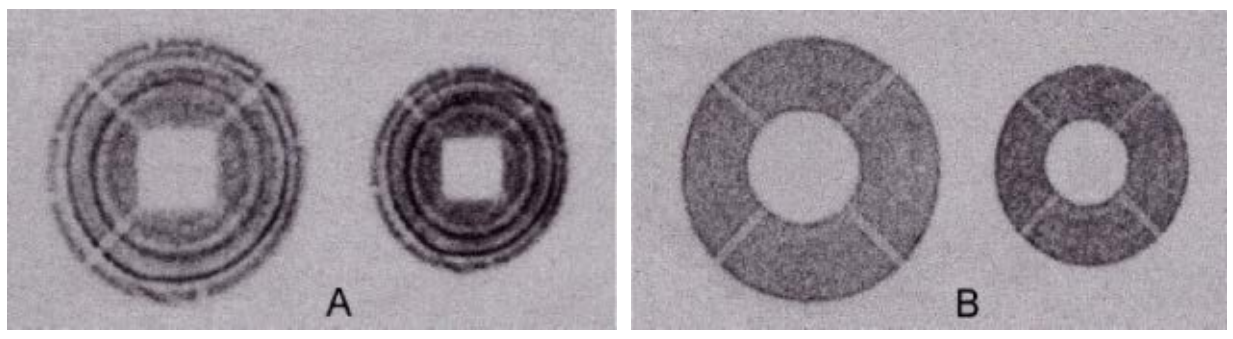

Fig. 6. Intra-focus patterns of a star obtained with a plate from zonal retouch (A) and stress figuring (B) (Franco-Belgium Schmidt Telescope - LOOM).

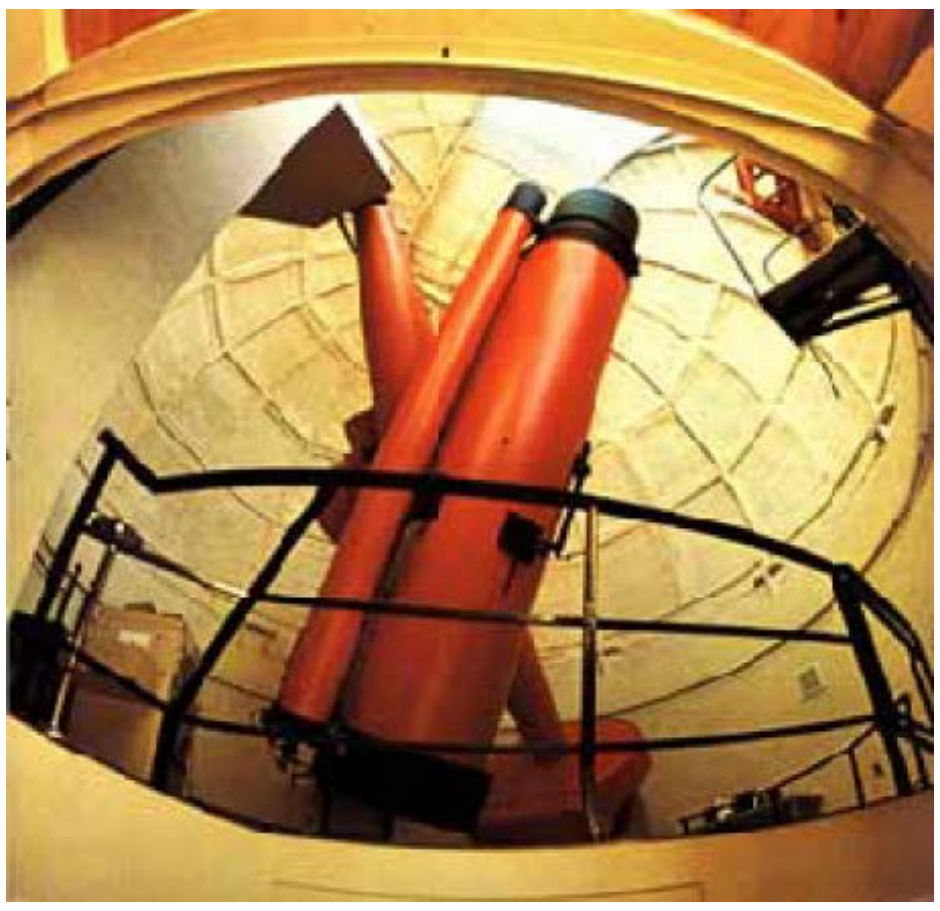

Fig. 7. Franco-Belgium 62/90-cm Schmidt Telescope at f/3.25 (Haute Provence Observatory & LOOM).

A UBK7 62-cm clear aperture corrector plate for the OHP Franco-Belgium Schmidt Telescope was aspherized by use of the *double-zone method* (Fig. 7).

## 5. Variable Curvature Mirrors (VCMs) for Telescope Arrays and Fourier Interferometers $z = A_{20}\, r^2$

Infrared Fourier transform interferometers and telescope array interferometers require use of particular optical designs as cat's eye systems (Fig. 8). Such retro-reflective systems are mounted on carriage translators to provide optical path variations. The field compensation or field cophasing is conveniently achieved by a variable curvature mirror (VCM).

This mirror—mounted at the focal plane of the cat's eye—generates a pure curvature mode, called Cv1 mode, of the form $z = A_{20}\, r^2$. Elasticity theory has provided highly deformable mirror satisfying the quarterwave criterion by use of metal alloys behaving linear stress-strain relationships.

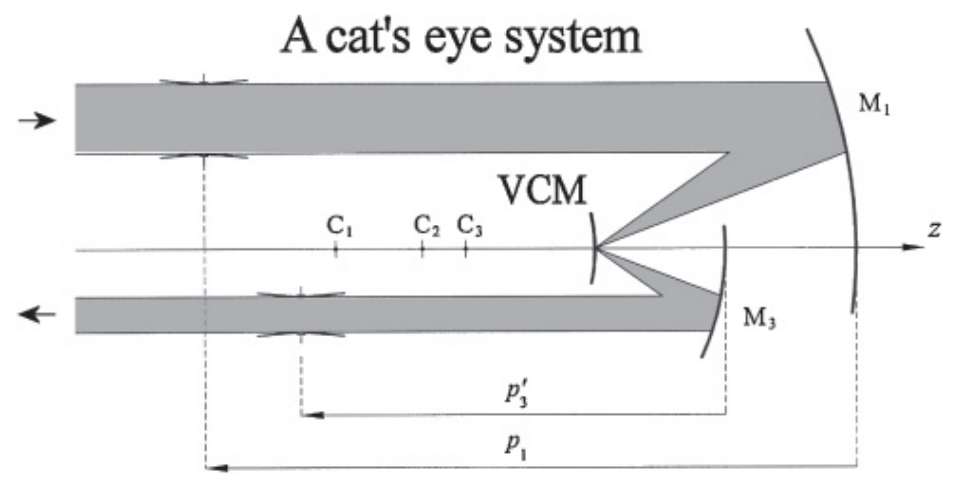

Fig. 8. Optical design of a cat's eye system with VCM.

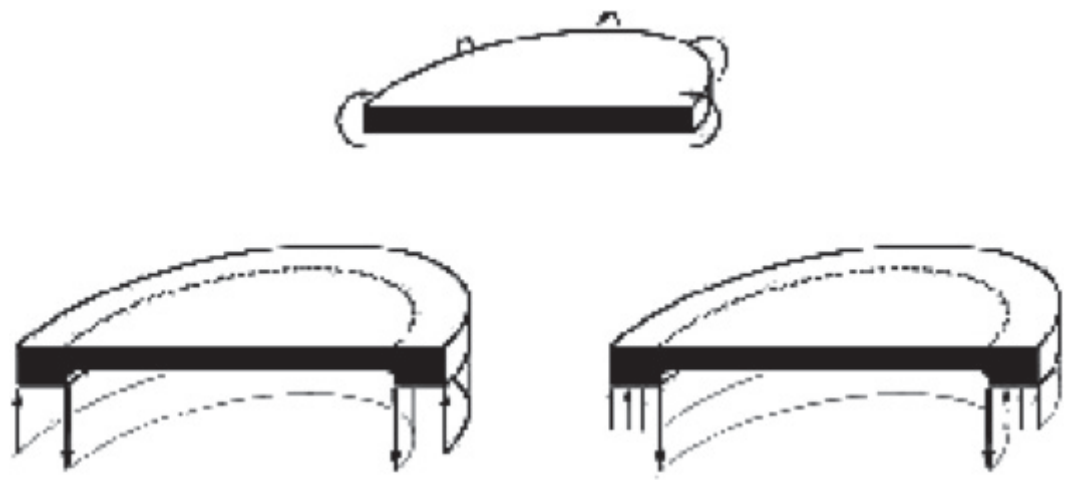

Fig. 9. VCM designs. Constant-thickness CTD class.

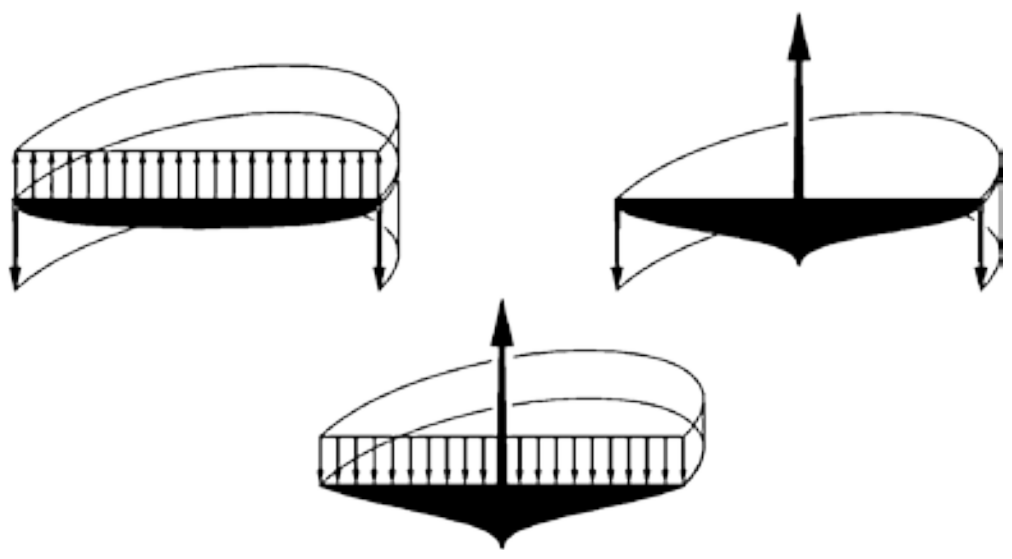

Fig. 10. VCM designs Variable thickness VTD class. Thickness distributions : $t/t_0 = (1-\rho^2)^{1/3}$, $(-\ln \rho^2)^{1/3}$, $(\rho^2 - \ln \rho^2 - 1)^{1/3}$.

Load configurations with constant thickness distributions – CTD class – of the substrate associated to a uniform perimeter bending moment are possible but would entail some practical complexity (Fig. 9).

More practicable loading configurations requiring use of variable thickness distributions—VTD class—have been found by Lemaitre 1976 [4] and developed at LOOM. Tulip form VCMs associated with a central force and cycloid like form VCMs associated with a uniform load provided useful devices (Fig. 10).

The thin plate elasticity theory provide CTD or VTD configurations that generates *Cv*1 modes of diffraction limited quality for zoom ranges extending, say, from f/∞ to f/5. For obtaining larger zoom ranges the elasticity theory of large deformations has been applied by Ferrari[5,6] to the design of cycloid-like form and tulip form VCMs. The two forms shown at top of Fig. 10, with load reaction at the edge, were built in stainless steel metal substrate by selecting the quenched alloy FeCr13 which exhibits a highly linear stress-strain relationship. The mirror design parameters are 16mm aperture, 300 μm central thickness and maximum flexural sag 400 μm at 7 bars loading typically. The reflective zone is simply supported by an outer rigid ring made in a single piece with the mirror. Interferometric tests were carried out with respect to six spherical calibers (Figs. 11–13).

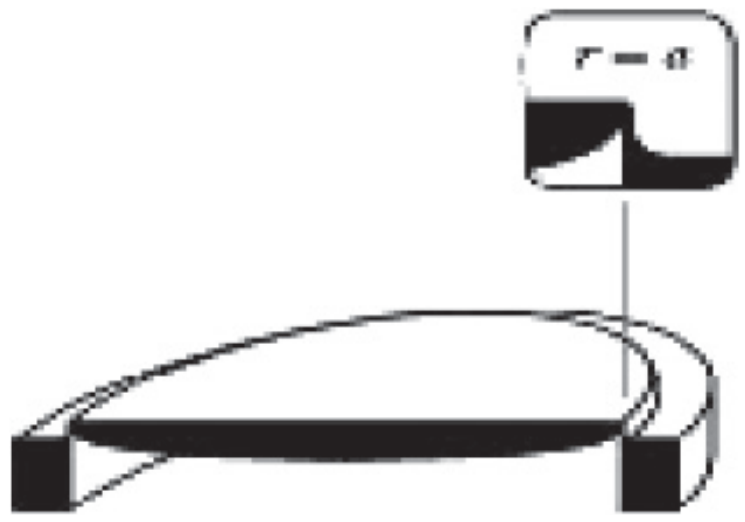

Fig. 11. Holosteric design of a cycloid-like VCM.

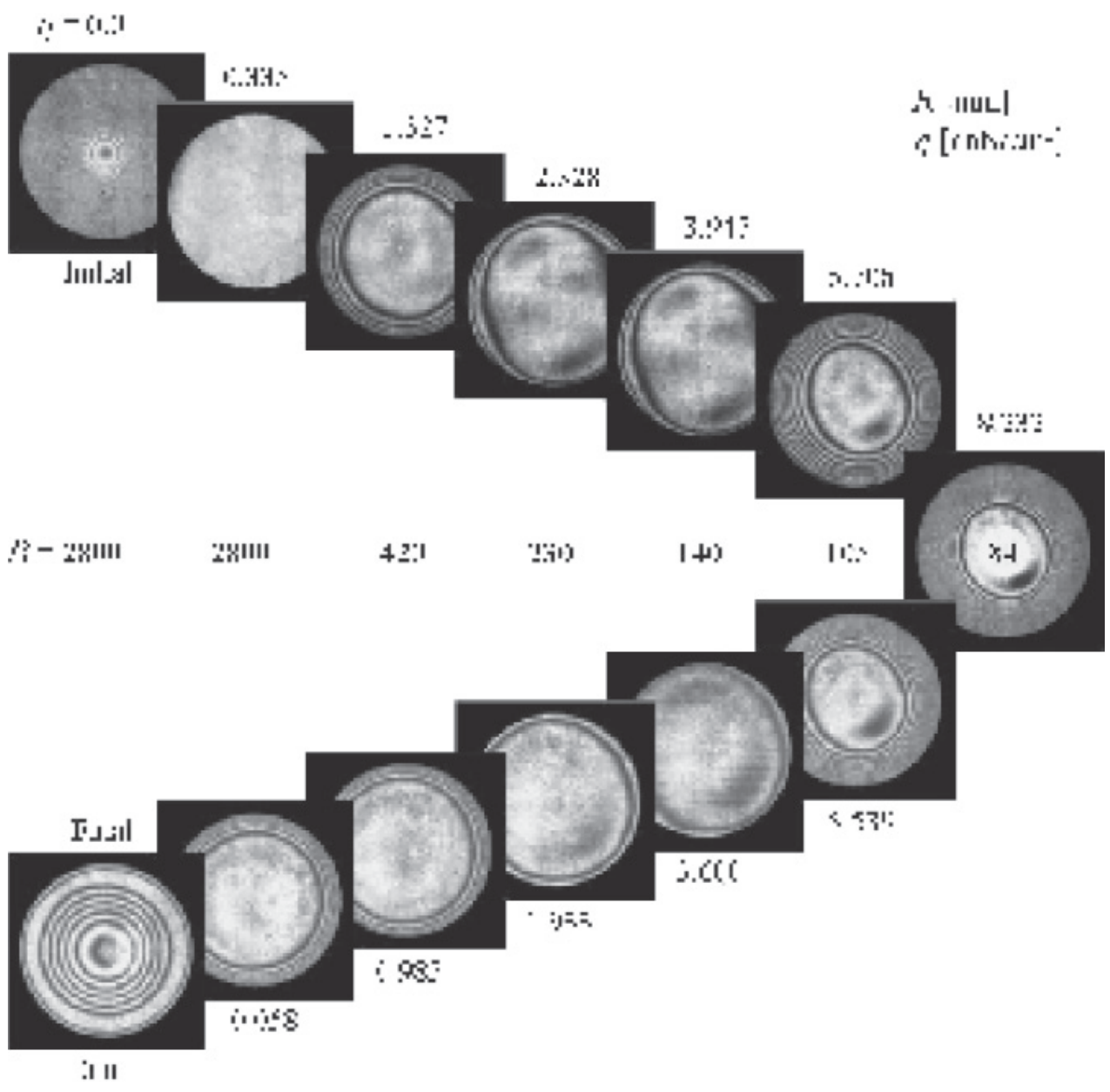

Fig. 12. VCM null tests of prestressing cycle for stress-strain linearization.[1,7] Calibers of radius of curvature R versus bending air pressure q. Typical maximum loading is 8 bars (LOOM).

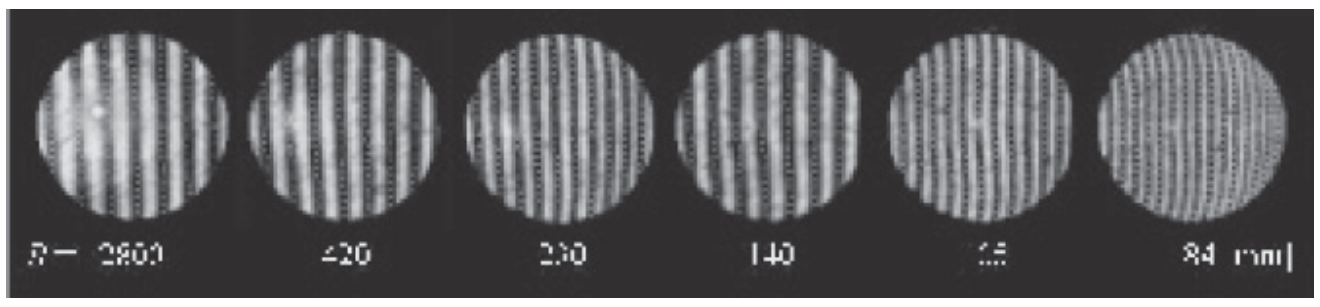

Fig. 13. Final tests after polishing of one of 16 VCMs for the ESO VLTI Array (LOOM).

The on-axis optical path cophasing of the VLTI Array uses translating delay lines by Derie et al.[8] that compensate for the path variation caused by diurnal rotation. Each delay line is equipped of a VCM at the focal plane of a cassegrainian cat's system for field cophasing (Figs. 14 and 15).

## 6. Mirrors Generating Single Aberration Modes with a Minimum Number of Actuators

Four geometrical classes for mirror substrates able to generate separately the single primary aberration modes *Sphe*3, *Coma*3, and *Astm*3 were investigated by Lemaitre[1,9] using a maximum of 2 or 3 actuators. These designs are derived from the various classes : *constant* or *quasi-constant thickness distribution* (CTD, q-CTD), *variable thickness distribution* (VTD) and a combination of them we called *hybrid thickness distribution* (HTD).

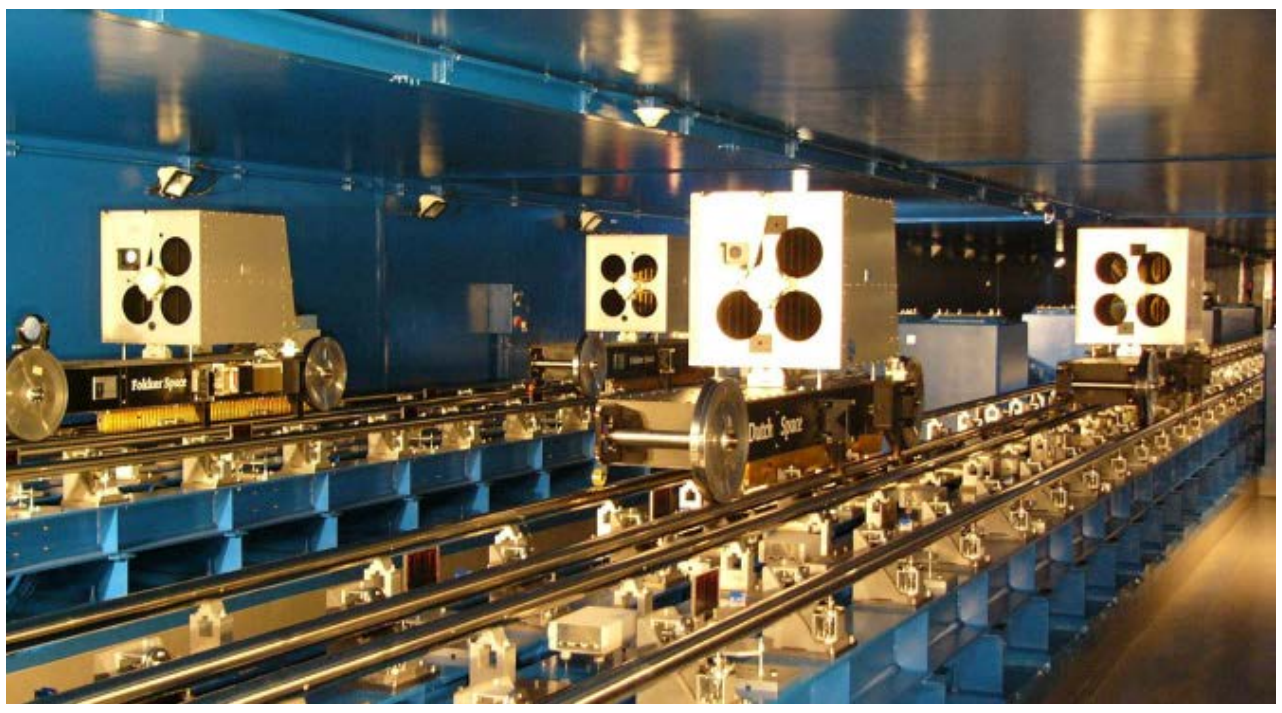

Fig. 14. Delay lines and cat's eye at VLTI tunnel (ESO).

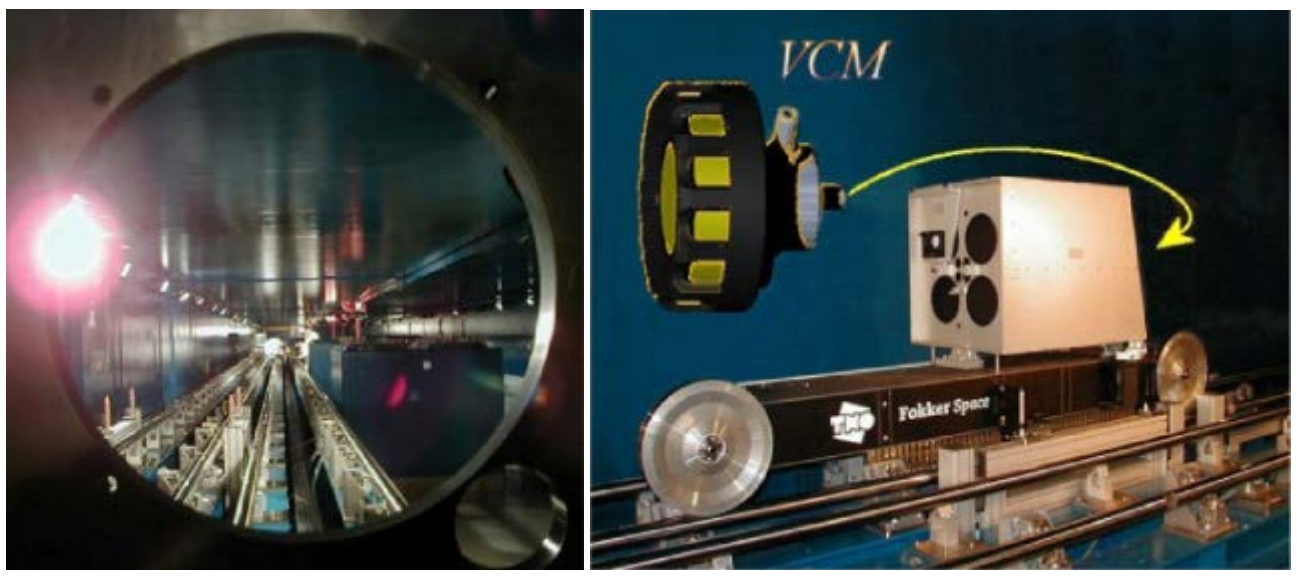

Fig. 15. VCM at carriage cat's eye focus (LOOM and VLTI). The implementation of VCMs on ESO-VLTI is i) 8 VCMs on 8 delay lines (4 UTs 8m + 4 ATs 2m) and ii) 8 VCMs on 4 ATs for the phase reference imaging combiner PRIMA.

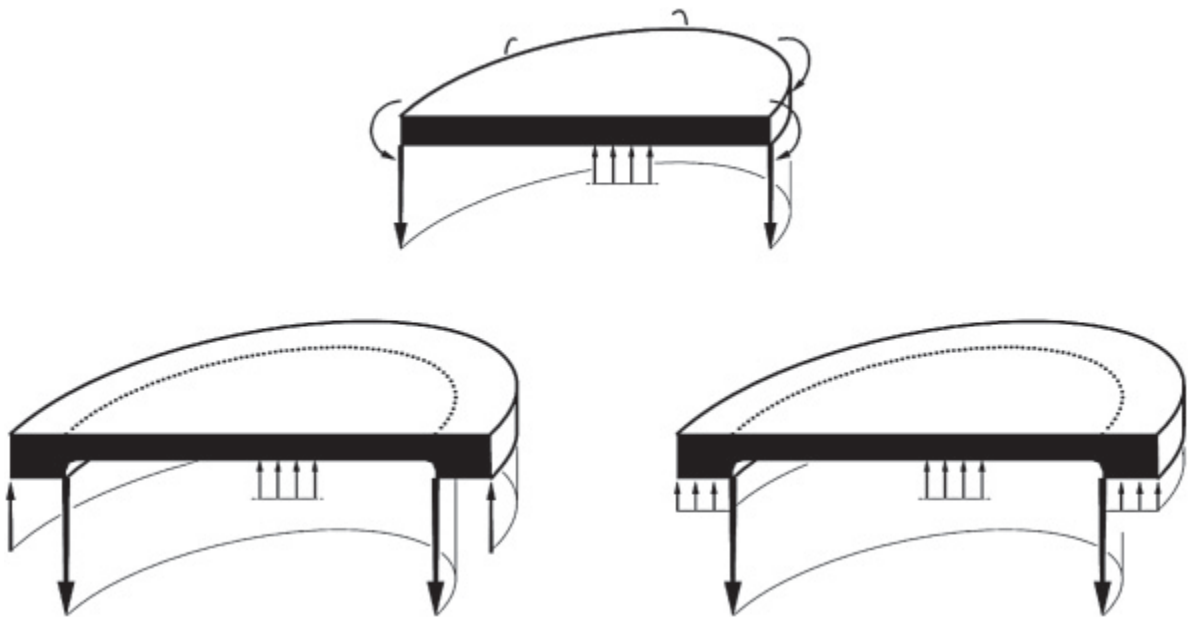

Fig. 16. Generating *Sphe*3-mode in the CTD class.

### 6.1 *Primary spherical aberration mode* $z = A_{40}\, r^4$

Several configurations were found and developed for mirrors generating a spherical aberration mode *Sphe*3, i.e., of the form $z = A_{40}\, r^4$. Similarly to the case of *Cv*1 mode, the CTD class provides configurations which lead to some difficulty in practice (Fig. 16) whilst designs in the VTD class are much easier to develop (Fig. 17).

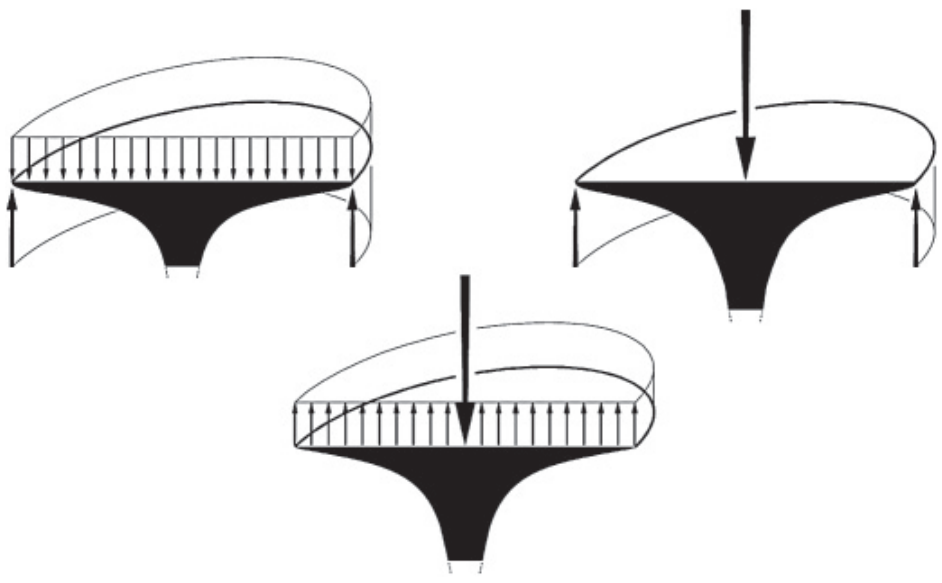

Fig. 17. Generating *Sphe*3-mode in the VTD class.
Thickness distributions:
Up-left $t / t_0 = [\rho^{-8/(3+\nu)} - 1]^{1/3}$. Up-right $t / t_0 = [\rho^{-8/(3+\nu)} - \rho^2]^{1/3}$
where ν is the Poisson ratio.

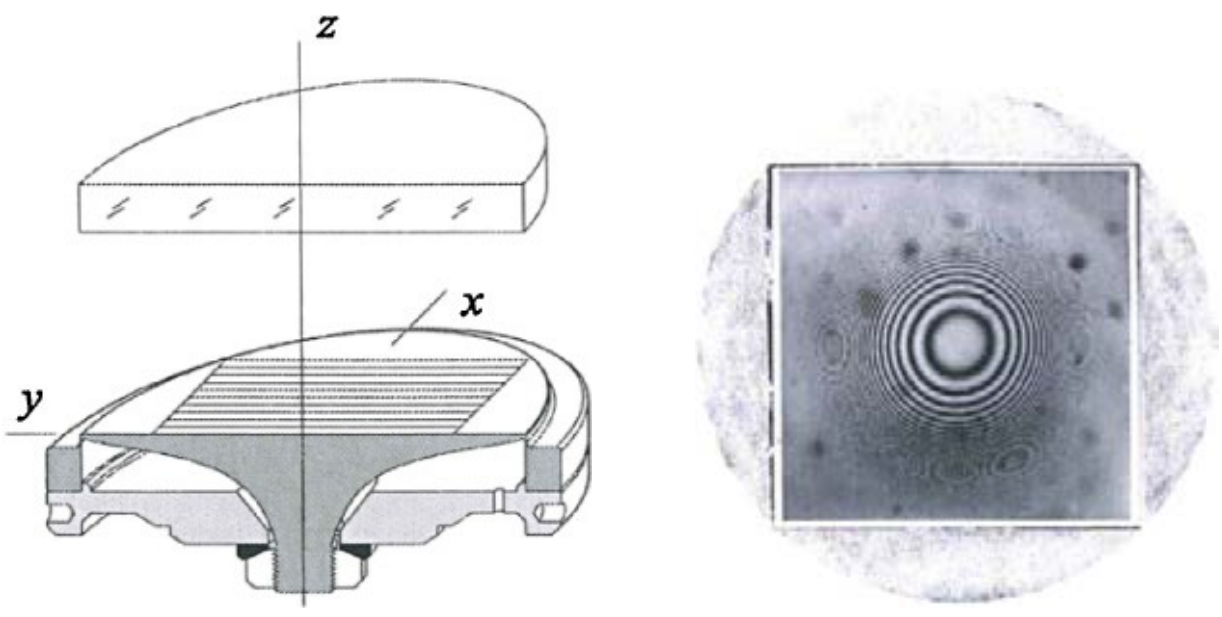

Fig. 18. Sphe3-corrective grating on Zerodur substrate from 2-stage replication (LOOM & Jobin Yvon Horiba).

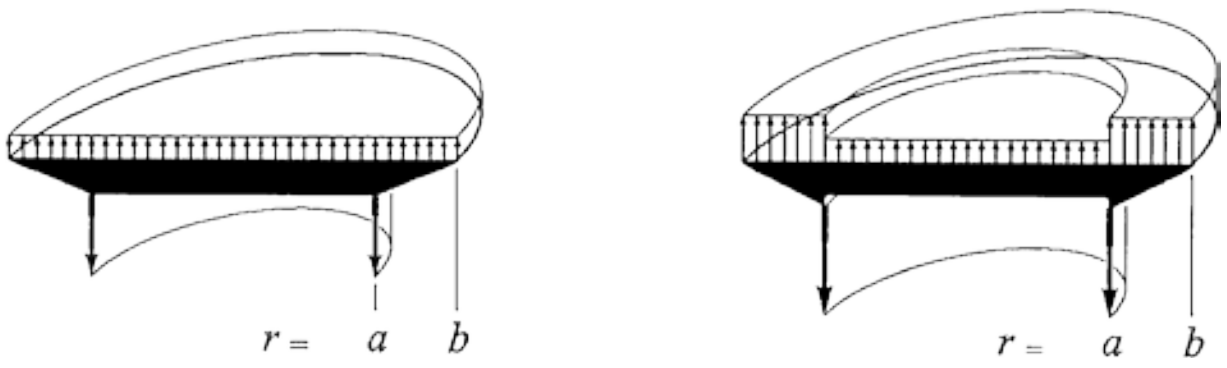

Fig. 19. Generating *Sphe*3-mode in the HTD class.

producing *Sphe*3 correcting diffraction gratings: A plane grating is first deposited on an intermediate plane-deformable tulip-form submaster.[1)] The final grating replica is obtained during controlled stress of the tulip-like form metal submaster in FeCr13 alloy (Fig. 18). Another possibility to generate a *Sphe*3-mode is to combines the CTD and VTD classes, then forming a hybrid thickness distribution (HTD) [1)] class (Fig. 19).

However most applications of active optics for correcting *Sphe*3-mode are usually combined with a *Cv*1-mode (see hereafter).

### 6.2 *Combined spherical aberration* $z = A_{20}\, r^2 + A_{40}\, r^4$

Generating flexures expressed by a combination of *Cv*1 and *Sphe*3 modes with $A_{20}A_{40} < 0$ allow reducing the stress level in the deformable substrate. Such profiles are of particular

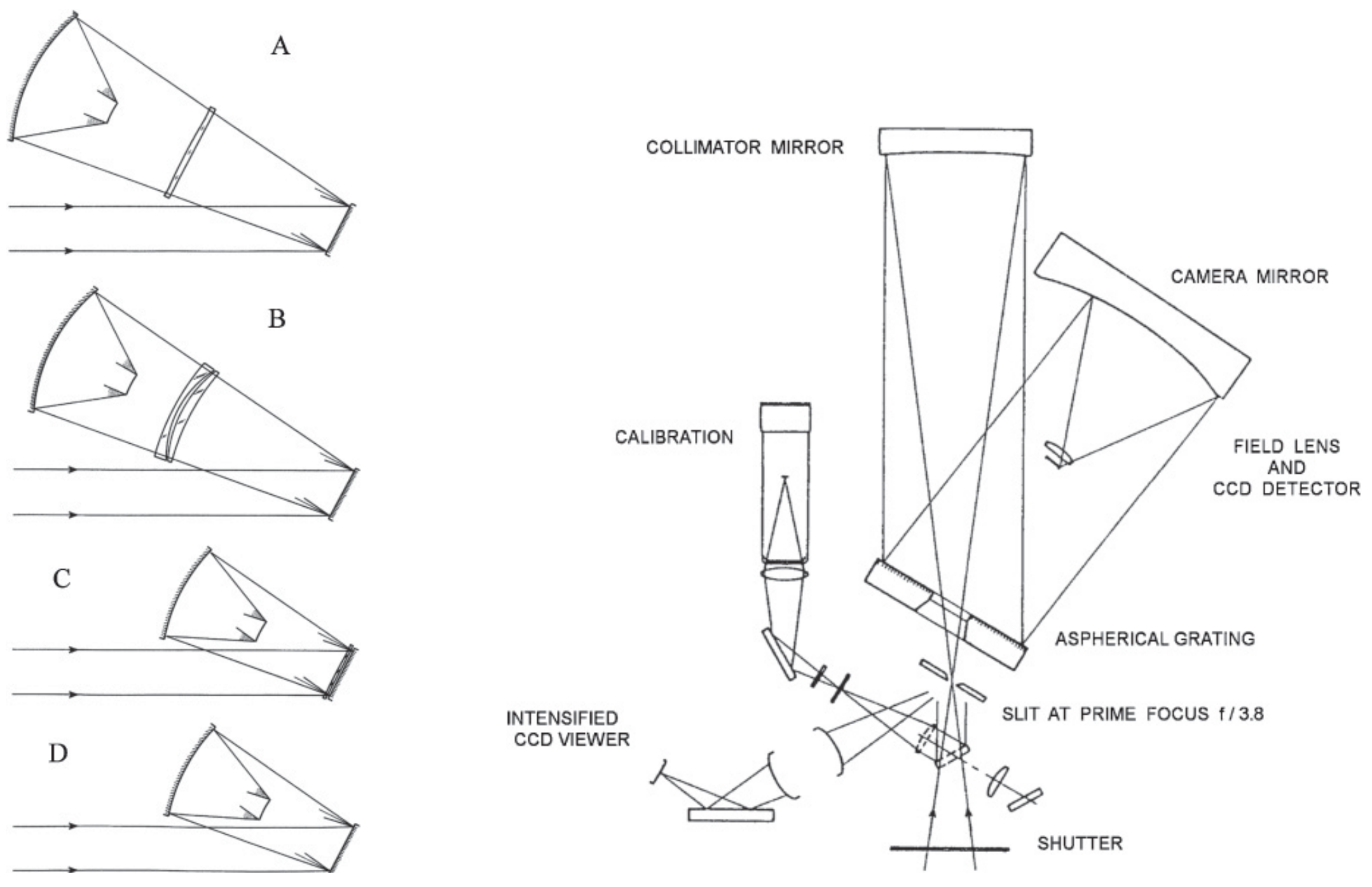

Fig. 20. (*Left*) Spectrograph design: Comparison of four camera optics with reflective gratings. Design (C) is with a double pass corrector plate at the grating whilst (D) is with an aspheric grating. (*Right*) Optical design of aspherized grating UV Prime Focus Spectrograph (CFHT).[1)]

interest for any reflective Schmidt systems as, for instance, high-throughput spectrograph types equipped with aspheric reflective gratings. One presents hereafter several applications and results.

### *6.2.1 Reflective Schmidt systems: Telescopes and spectrographs*

An important law for the design of a corrector of any reflective Schmidt system was found by Lemaitre:[1,10)] *A reflective surface where the radial variation of its local curvature is perfectly balanced minimizes the field aberrations of any reflective Schmidt system*. This applies to a Schmidt primary mirror—as giant Schmidt Lamost—as well as to an imager-spectrograph camera optics using a reflective corrector or an aspherized grating.

Aspherized reflective gratings provide a very compact optical design of the camera optics (Fig. 20). Deformable submasters with built-in boundary at the edge and quasi-constant thickness distribution (q-CTD) were developed at LOOM to generate aspherized gratings through the double replication technique (Fig. 21). Many spectrographs have benefited of these gratings (CFHT Mauna Kea, Pic du Midi, Haute Provence and Purple Mountain Observatories, Space Mission Odin-Osiris, etc). These gratings were built either for on- or off-axis mountings (Fig. 22).[1)]

### *6.2.2 Asphérization of a thin shell secondary mirror*

The hybrid class (HTD) allowed stress figuring for the hyperbolization a secondary mirror, 1.2m in diameter, as thin shell adaptive secondary mirror for one of the ESO-VLT 8m-Unit (Fig. 23). The process has been proposed and developed by Hugot et al.[1]

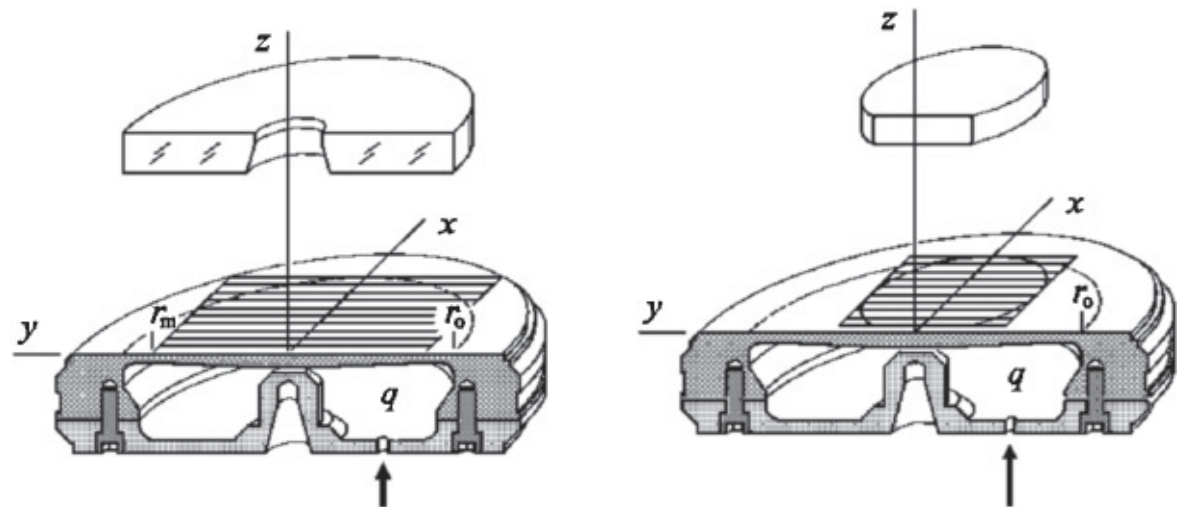

Fig. 21. Deformable submaster of quasi-constant thickness distribution (q-CTD) for grating aspherizations.

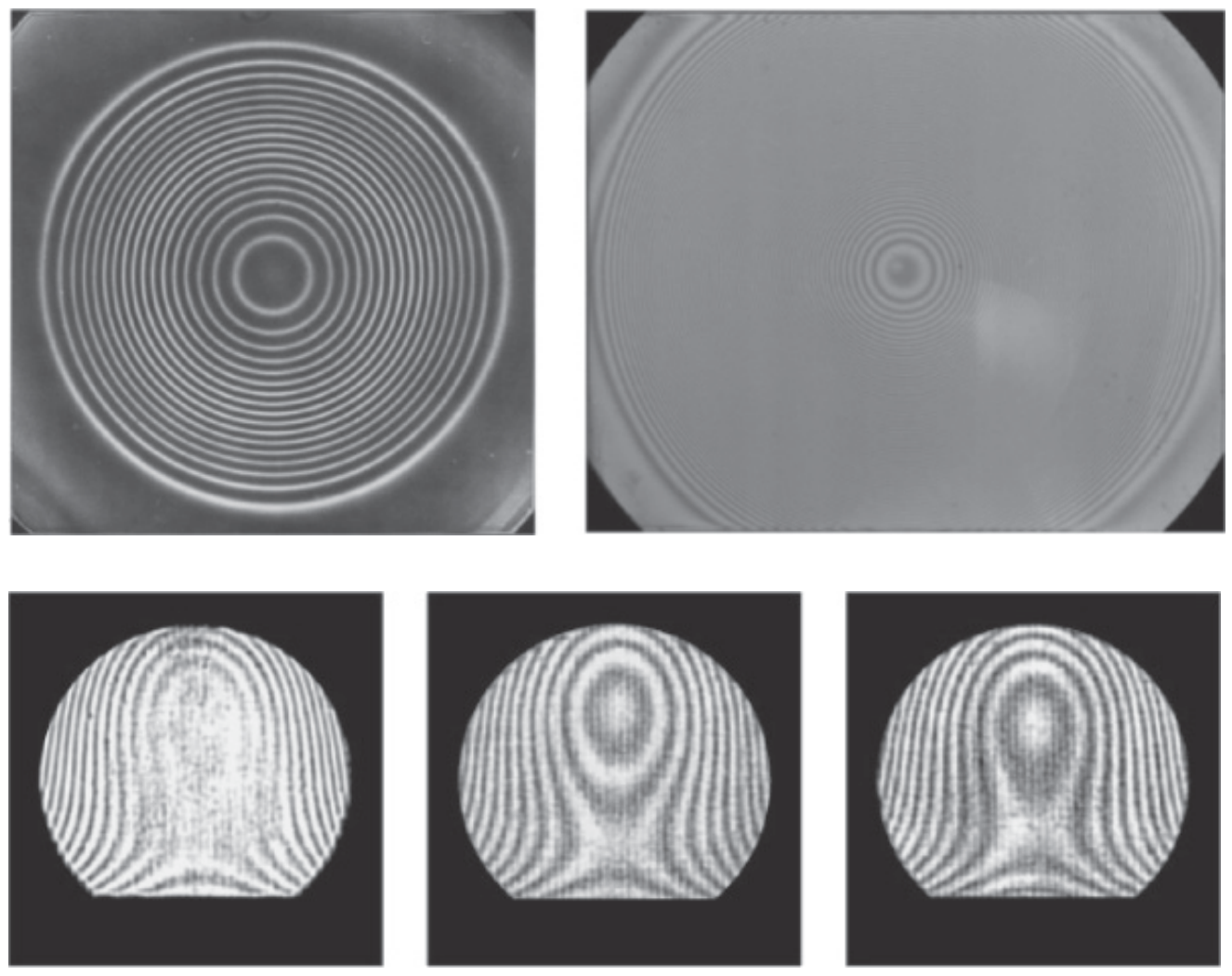

Fig. 22. He–Ne fringes with respect to flat of on- and off-axis reflective grating replicas aspherized on Zerodur substrate. MARLY, CARELEC and ISARD spectrographs (LOOM).

*6.3 Primary coma mode* $z = A_{31}\, r^3 \cos\theta$

Analytic investigations with the thin plate theory of circular plates show that primary coma, *Coma*3, can be generated from either, in the CTD class, by use of a bending moment whose lineic distribution along the substrate contour is of the form cos$\theta$, or, in the VTD class, by a tulip form on which a lineic net shearing force at the contour varies as cos$\theta$ in reaction to a bending moment at the substrate center (Fig. 24). An easily practicable configuration is a *tulip form* where a one-directional bridge is built-in at the central zone and opposite forces are applied to the bridge ends on a simply-supported outer cylinder. A *hybrid form* (HTD) is an equivalent alternative solution (Fig. 24).[1)]

Applications were developed for obtaining a low dispersion coma-corrected transmission grating. This design was used for slitless spectroscopy in convergent beam at the f/8 Cassegrain focus of CFHT (Fig. 25). A plane grating 75-l/mm was first deposited on a CTD deformable submaster (Fig. 26), and then replicated during stressing onto a transmission substrate.[1)]

*6.4 Primary astigmatism mode* $z = A_{22}\, r^2 \cos 2\theta$

Primary astigmatism mode, *Astm*3, can be generated either from CTDs or VTDs. The flexure to be achieved is a saddle-like surface i.e., with opposite curvatures in main orthogonal directions. In the CTD class, a basic solution is with bending moments distributed in cos2$\theta$ along the circular contour. With VTD class the basic solution is a *cycloid-like* thickness on which a net shearing force of the form cos2$\theta$ is applied along the mirror contour (Fig. 27). Practicable configurations were developed at LOOM by use of either four outer bending bridges for CTDs or opposite force-pairs on simply-supported outer ring fo VTDs.[1)]

Applications were developed for the optical concept of extreme-ultraviolet *single-surface spectrographs* CDS and UVCS of the Soho Mission—still continuing solar observations at Lagrange point L1 in 2012. These designs by Huber et al.[12)] use toroid reflective gratings which provide two stigmatic points in the dispersed field of the spectra (Fig. 28). The process for obtaining the gratings uses a first replication of a spherical grating onto a deformable submaster.[1)] Stainless steel submasters, designed in the VTD class, were designed with the cycloid-like form and four azimuthal bridges (Fig. 29). A second stage uses a four force stressing for the final replication on rigid Zerodur substrates.

Other applications have been found useful for the off-axis optics of planet finder coronagraph Sphere, a second generation of ESO-VLT instrumentation. Three mirrors

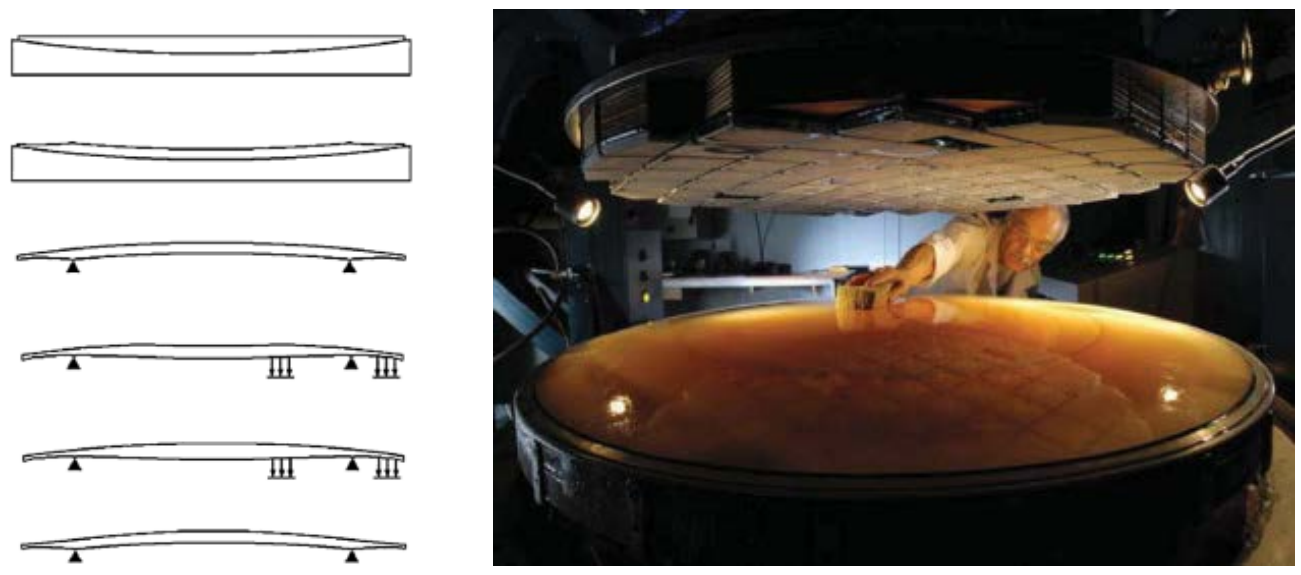

Fig. 23. Thin shell aspherization process for adaptive Cassegrain mirror of the ESO-VLT (LOOM).

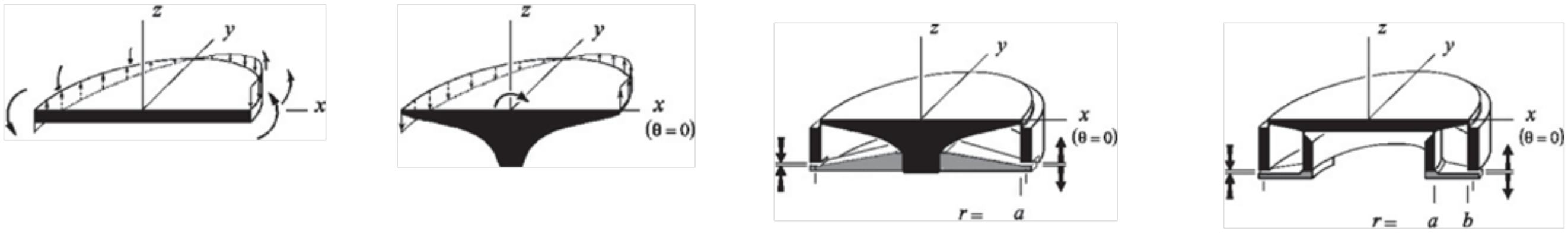

Fig. 24. *(Left*) Generating Coma3-mode with CTD and VTD. (*Right*) Practicable solutions with VTD and HTD.

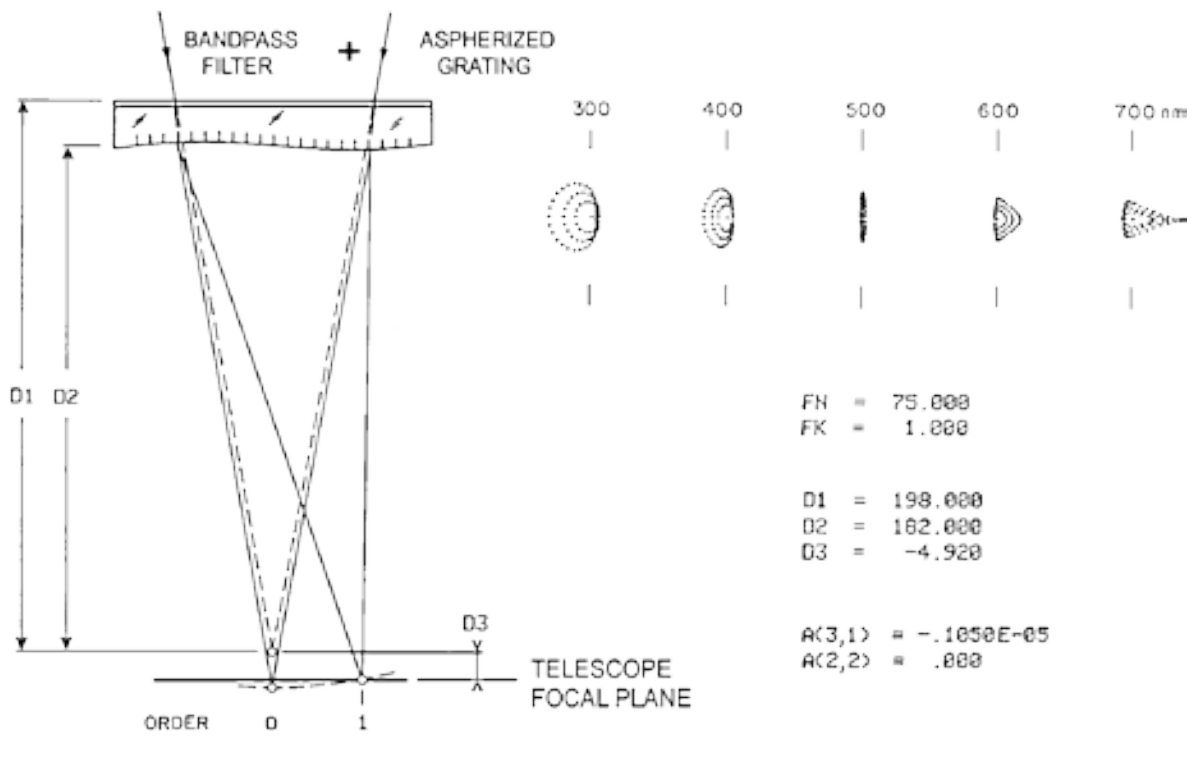

Fig. 25. Objective- or slitless-spectroscopy with *Coma*3 corrected transmission gratings at CFHT Cassegrain focus.[1)]

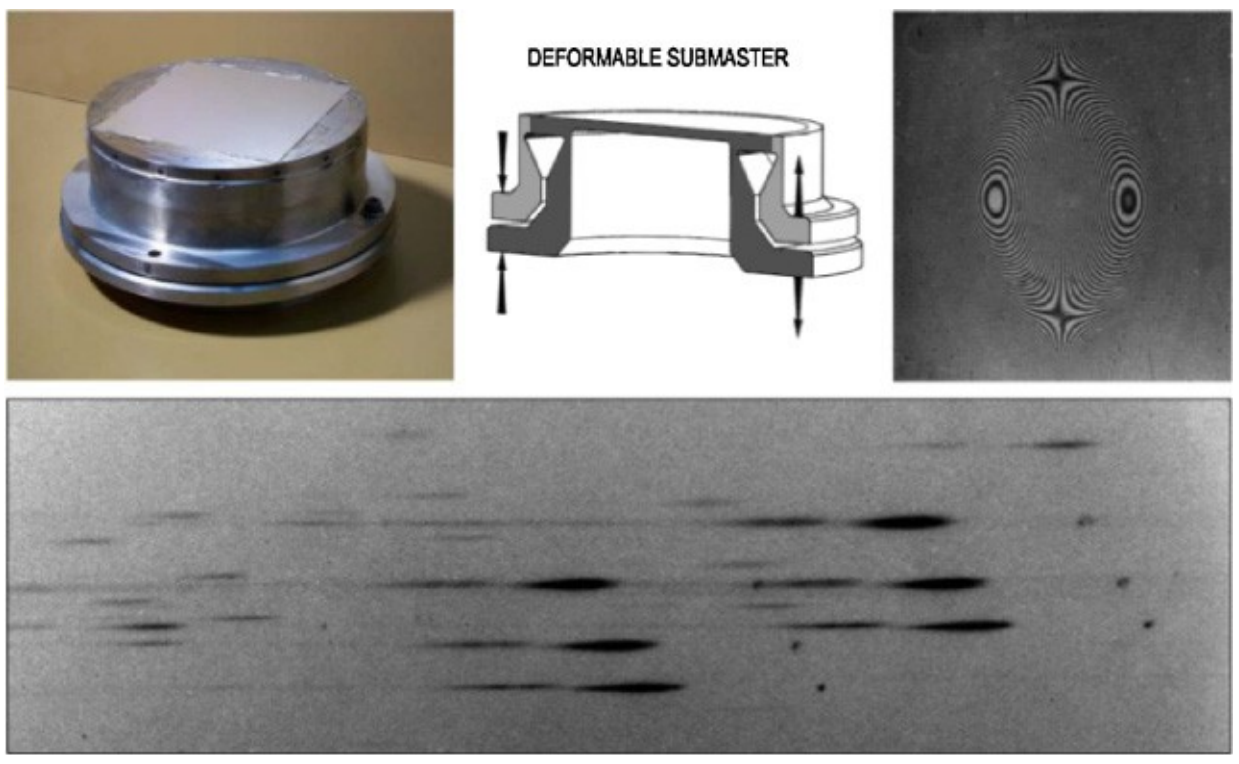

Fig. 26. Active submaster for *Coma*3-corrected grating working in telescope convergent beam (LOOM & Hyperfine Corp.).

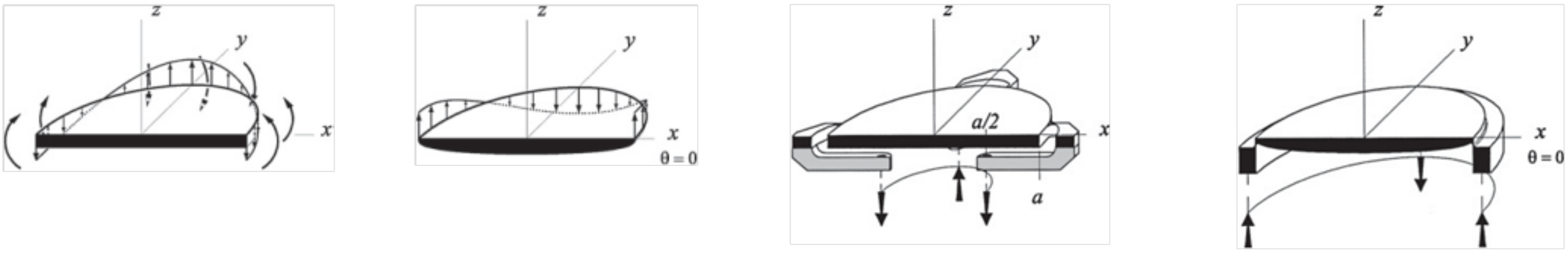

Fig. 27. (*Left*) Generating *Astm*3-mode with CTD class and VTD class $t / t_0 = (1 - \rho^2)^{1/3}$. (*Right*) Two practicable solutions.

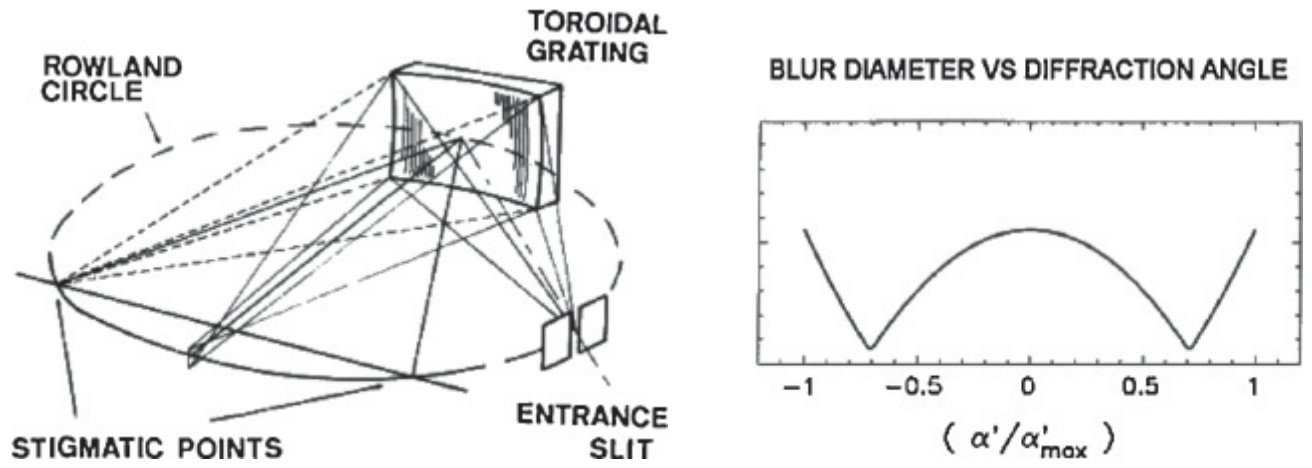

Fig. 28. Single surface CDS and UVCS Spectrographs of the ESA NASA Soho Mission. The optical design with a toric grating provides two stigmatic points (ETH-Zurich).

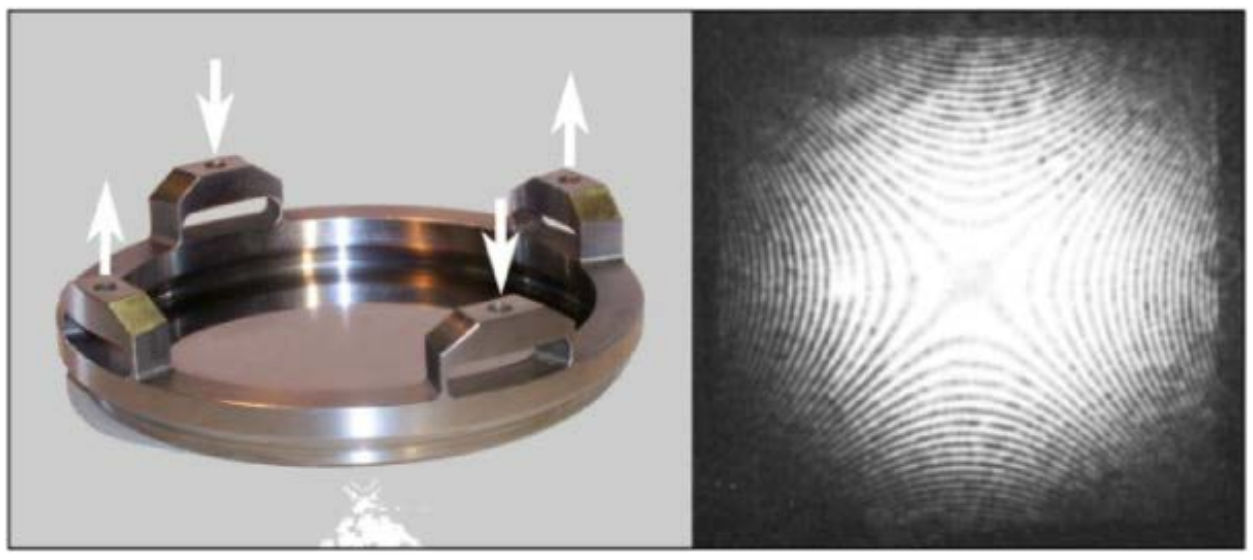

Fig. 29. Submaster and fringes of a toric *Astm*3-corrected grating replica of the Soho Mission (LOOM, ETH-Zurich & Bach Research Corp-Boulder).

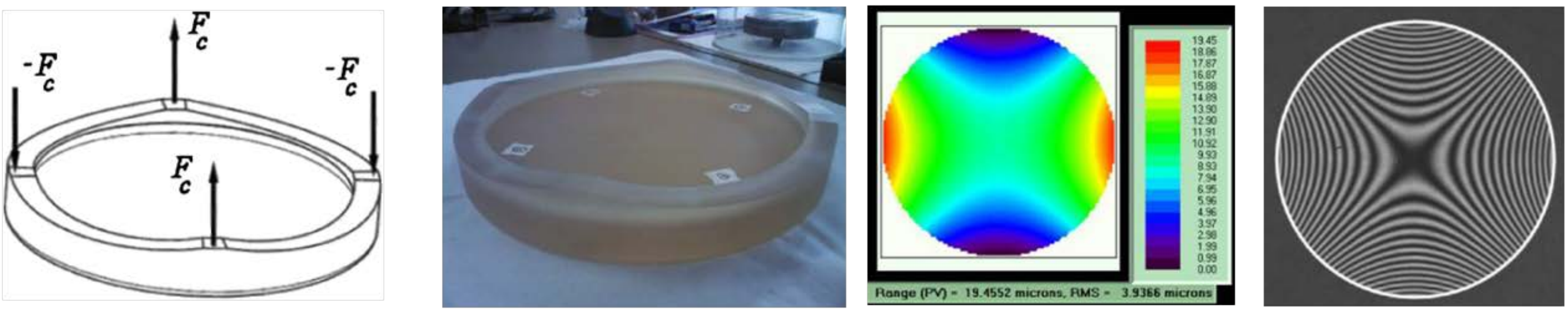

Fig. 30. Elasticity design of a 40 cm toroid mirror with azimuth-modulated thickness edge for SPHERE next generation planet finder of the ESO-VLT. This instrument includes three different toroid mirrors (LOOM).

differing in size were designed by Hugot et al.[13] as CTD vase form mirrors where the outer rings have azimuthal modulated axial thicknesses in between the four bending forces (Fig. 30). Results show that all mirrors have a deviation error to theoretical toroid shape smaller than 15 nm rms.

## 7. Aspherization of Axisymmetric Mirrors with Fast f-Ratios: Stress Figuring or In Situ Stressing?

If the mirror curvatures is faster than, say, f/3 then active optics aspherization requires taking into account the stresses induced at the middle surface of the substrate. The shallow shell theory takes these stresses into account for the axisymmetric case (Fig. 31) and allows solving analytically the problem of determining the radial thickness distribution for generating the required asphericity—e.g., a conicoid or a spheroid optical shape—from a spherical figure bent by uniform loading. This theory satisfactory relies on the strong dependence of the flexure with respect to edge boundary.

The three basic studied configurations are a *vase form* — which implicitly contains *a menis-cus form* by removing the outer ring—and a *closed form* (Fig. 32). Use of the shallow shell theory on $N$ successive ring segments linked together allow us to determine via iterative process the radial thick-ness distribution set $\{ t_n \}$ of the mirror.[1]

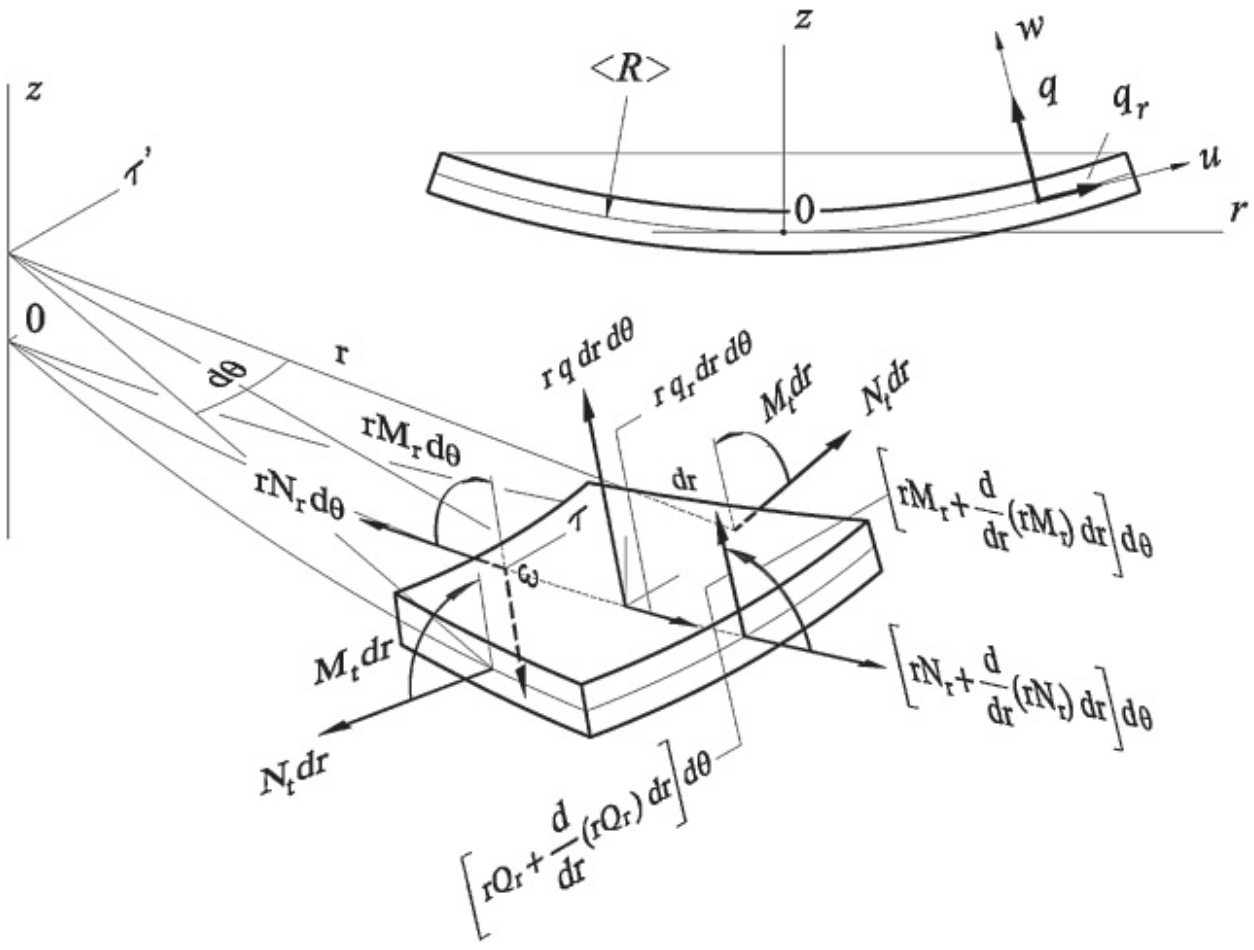

Fig. 31. Shallow shell theory—Equilibrium of elementary forces including those at the middle surface.

*7.1 In-situ parabolization of concave mirrors*

The parabolization of a concave mirror by stress figuring is a most difficult task because the uniform load to exert is not a partial vacuum—which would naturally flatten the mirror against the table of the machine—but an inner air pressure which then requires use of a continuous and accurate reaction system along the mirror contour. Depict of this tremendous difficulty for large telescope mirrors, a more natural alternative for smaller size mirrors, say up to 4m diameter, is to make a spherical figuring without stress and then to practice in-situ *partial vacuum at the telescope* by closing the rear side of the mirror.

Investigation in the case of a holed mirror aspherized by in situ stressing led us to retain a *vase shell* design with an optimized radial thickness distribution $\{t_n\}$ for an f/1.75 paraboloid 20 cm clear aperture mirror in Zerodur. The back side of the outer ring is given a folded L-shape to increase the perimeter rigidity without overweighting (Fig. 33). The central hole is closed with a paste on a soft sliding tube.

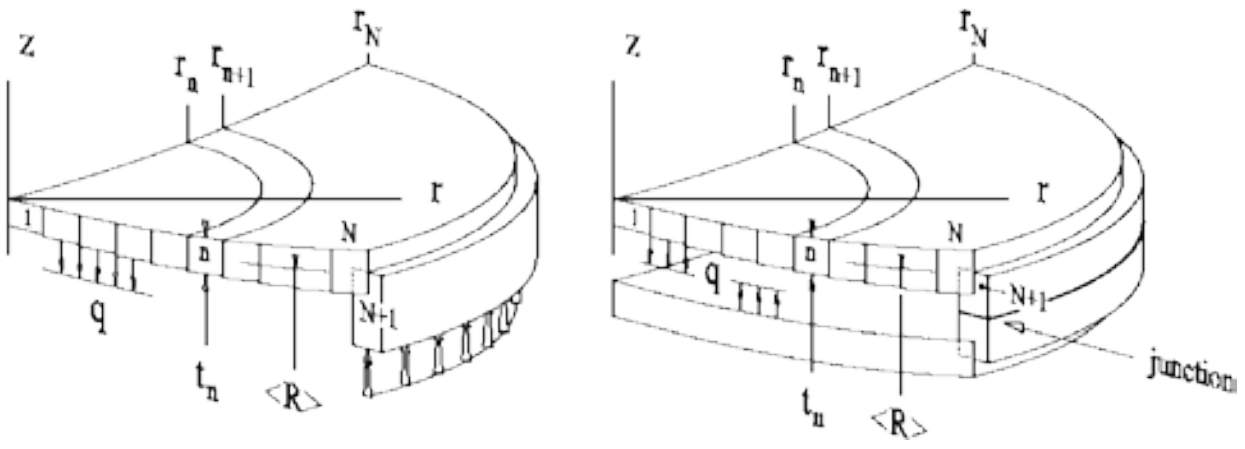

Fig. 32. Geometries of a *vase form* and *closed form*.

Another inner tube applies, via a conical spring, a shearing force which provides an equivalent distribution to that of plain mirror.[1)]

*7.2 Mirror hyperbolizations of a modified-Rumsey three-reflection telescope*

A very interesting optical system for wide-field skysurveys in astronomy—typically 1$^o$ or 2$^o$ FoV—is a threereflection modified-Rumsey telescope (Fig.34). This *flat field anastigmatic system* avoids use of 3-lens correctors of two mirror systems or 2-lens correctors for flat-fielded Ritchey Chrétien systems, and then is completely free from chromatic aberrations except that due to spectral band filter and detector window plate as for all usual optical systems. In a modified-Rumsey form, the elasticity analysis shows that both primary and tertiary mirrors can be in-situ hyperbolized simultaneously with a large single spherical tool on a whole common substrate which is designed in a *double-vase form*.[1,14)]

Two 50 cm aperture, 2_ FoV, modified-Rumsey telescope demonstrators, MiniTrust I and II, were built at LOOM. All three mirrors were given a spherical figure with only two rigid laps. The *double vase substrate* of M1–M3 mirrors allowed their simultaneous hyperbolization by in-situ stressing. Hyperbolization of *tulip form* M2 was achieved by stress figuring (Fig. 35). Results from whole telescope autocollimation test show a deviation error smaller than 280 nm PtV, i.e., 48 nm RMS for a single-pass wavefront (Fig. 36).

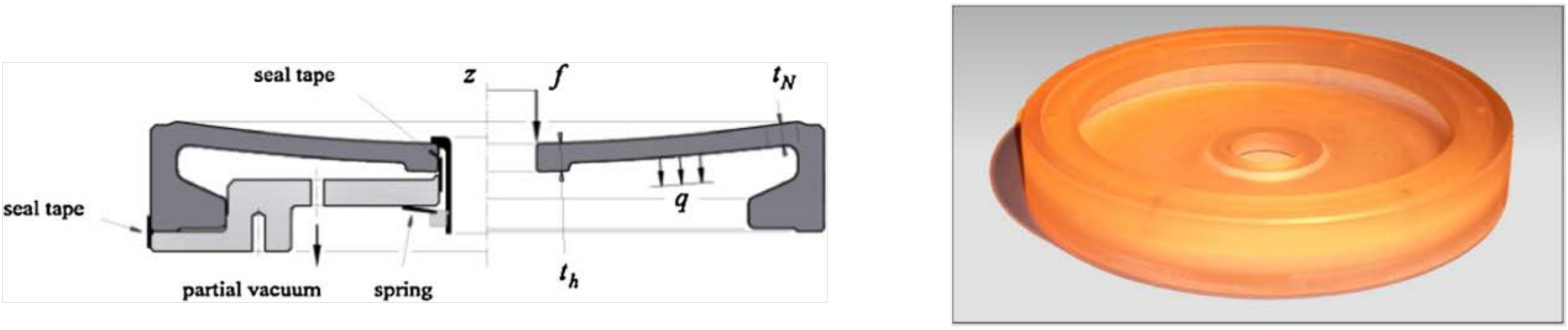

Fig. 33. Elasticity design and view of a holed f/1.75 vase form mirror parabolized by in-situ stressing (LOOM).

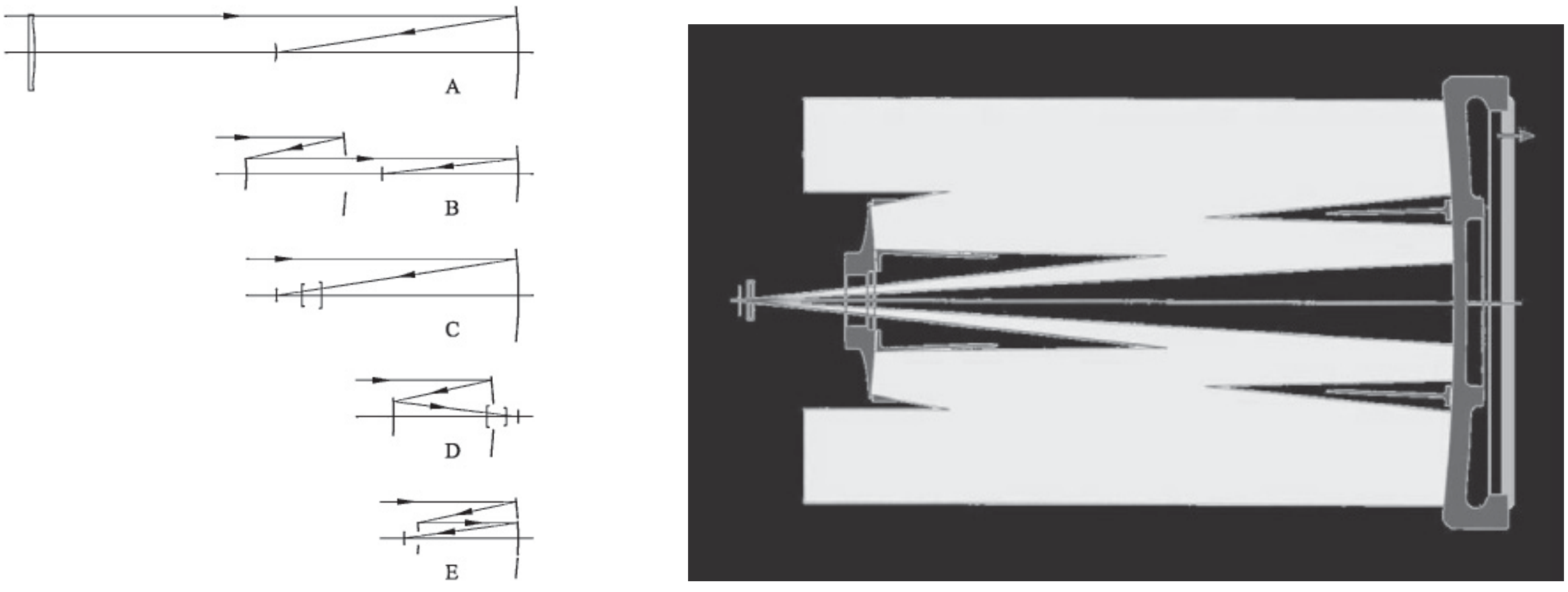

Fig. 34. (*Left*) Comparison of five wide-field telescope designs. (*Right*) Three-reflection telescope MiniTrust: Modified-Rumsey flat-field anastigmat design (E) *with slope and curvature continuity* between M1 and M3 mirrors.

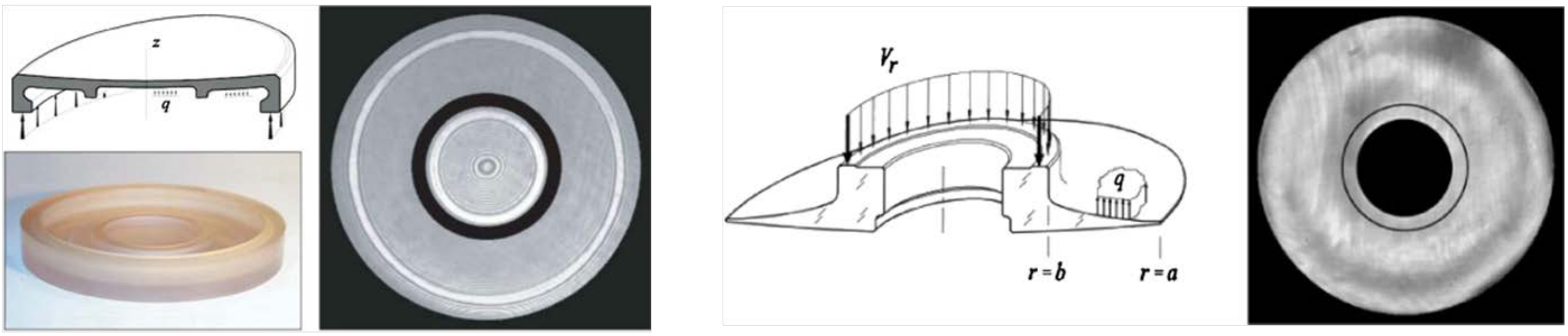

Fig. 35. MINITRUST mirrors. (*Left*) *Double-vase form* M1–M3 mirrors hyperbolized by in-situ stressing. (*Right*) Tulip form M2 mirror hyperbolized by stress figuring (LOOM).

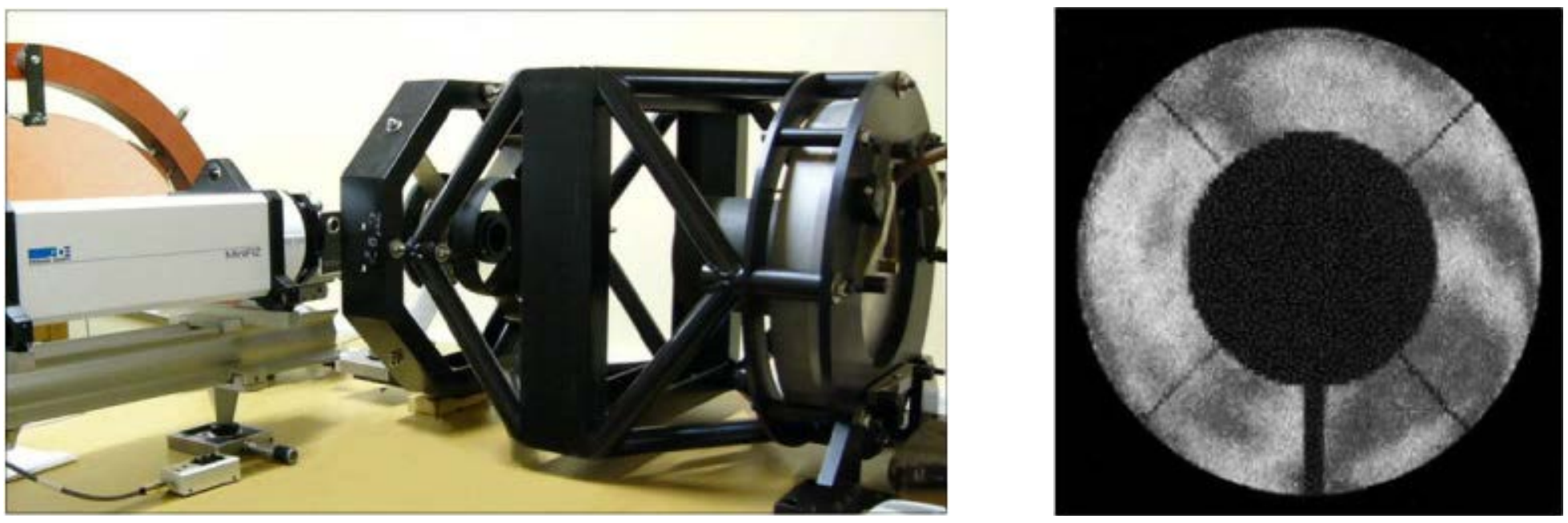

Fig. 36. MINITRUST view under optical testing. Wavefront error smaller than 48 nm-rms single-pass (LOOM).

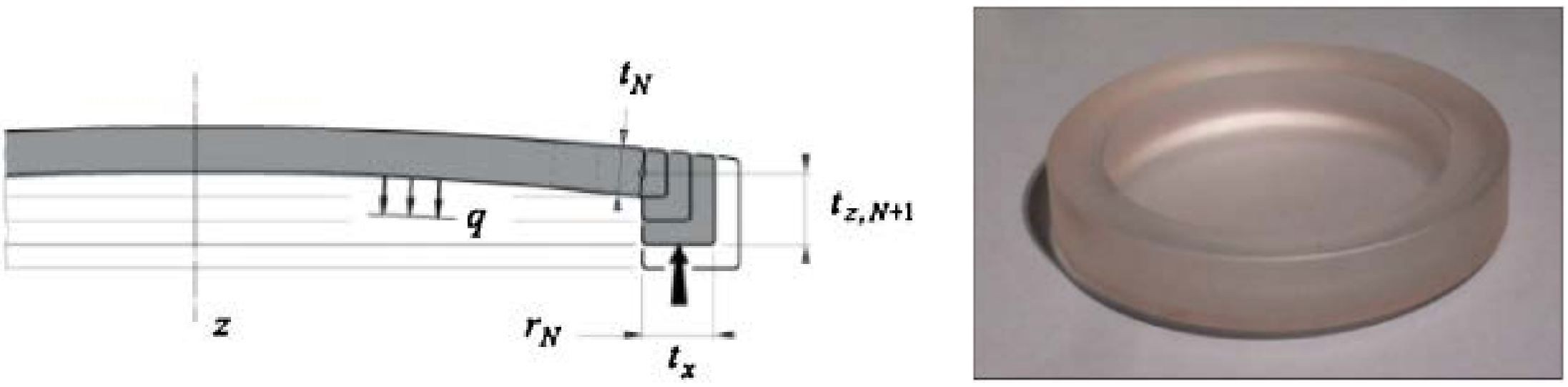

Fig. 37. Convex vase mirrors parabolized by stress figuring. Material: fused-silica ULE (LOOM).

### 7.3 *Aspherization of convex mirrors by stress figuring*

The aspherization of telescope secondary mirrors into a paraboloid or hyperboloid by uniform loading and stress figuring comes straightforwardly. This because partial vacuum is applied to the rear face of the mirror which then naturally flatten it on the table of the surfacing machine. We developed the parabolization of *vase form Cassegrain mirrors* made of ultralow expansion titanium-silica (ULE) for 1.5-m afocal telescopes. These mirrors were designed and built for Labeyrie[15)] at GI2T-CERGA to provide a beam compression-ratio of 20 (Fig. 37).

Two 25 cm aperture *vase form* convex mirrors in Zerodur were hyperbolized by stress figuring. For compactness, their design used an inner folded outer ring (Fig. 38). The thin thickness of the reflective area was useful to achieve a backside cooling.[1)] These are Cassegrain mirrors the 1m RC Themis Solar Telescope at Canary Islands (Fig. 39).

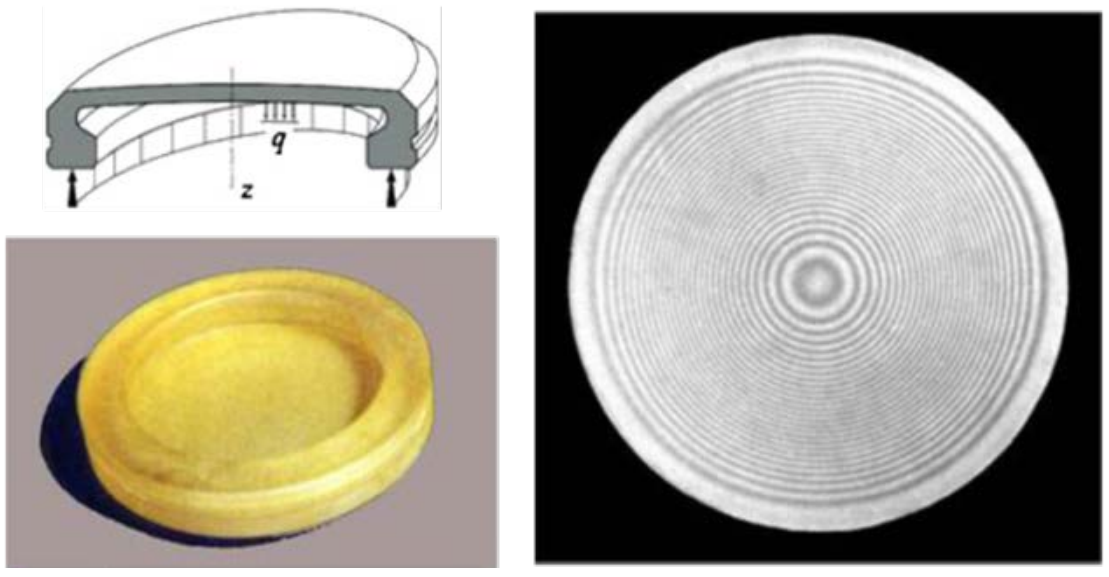

Fig. 38. Vase form secondary mirror of THEMIS Ritchey-Chrétien telescope hyperbolized by stress figuring (LOOM).

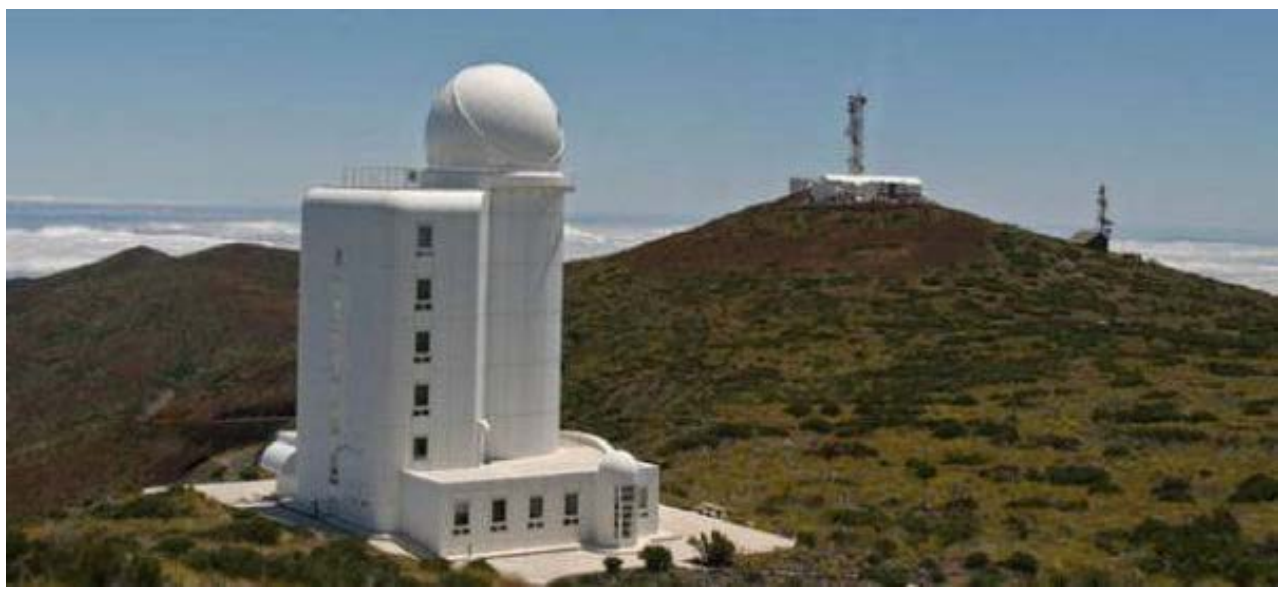

Fig. 39. France-Italy 1m Ritchey-Chrétien Solar Telescope THEMIS at Canary islands (INSU, NRC).

## 8. Multimode Deformable Mirrors (MDMs): Vase Form–Meniscus Form

Except for very special cases—as *Astm*3 and *Cv*1 modes which can be generated by a same *cycloid-like* thickness $t / t_0 = (1- \rho^2)^{1/3}$ in the VTD class with very few actuators—the superposition of various optical modes requires an elasticity design which belongs to the CTD class. Vase form or meniscus form *multimode deformable mirrors* (MDMs) can be bent by forces applied to outer radial arms and an optional uniform load all over the mirror surface.[1,16] For large telescope mirrors one prefers use of force actuators uniformly distributed over their rear surface.

### 8.1 MDMs with outer arms and Clebsch–Seidel modes

Because of the similar form of Seidel optical modes and Clebsch flexural mode that are general solutions of Poisson's bilaplacian equation $\Delta^2\Delta^2 z + q/D = 0$, where $q$ is the uniform load and $D$ the constant rigidity, one shows that from thin plate theory there exists a common subset of modes similar to Seidel modes. We called them Clebsch- Seidel modes.[1]

Denoting a mode as $z = A_{n,m} \cos m\theta$, this subset is generated by terms $m = 0$ for $q = constant$, and $m = n$ and $m = n - 2$ for $q = 0$. These terms are the *Sphe*3 mode and all modes of the two lower diagonal lines $D_1$, $D_2$ of the optics triangle matrix modes (Fig. 40). Shown interferograms were obtained from a 20 cm MDM with $K = 12$ arms for which axial forces $F_{a,k}$ and $F_{c,k}$ are applied to the inner and outer end of arm-$k$ respectively (Fig. 41).

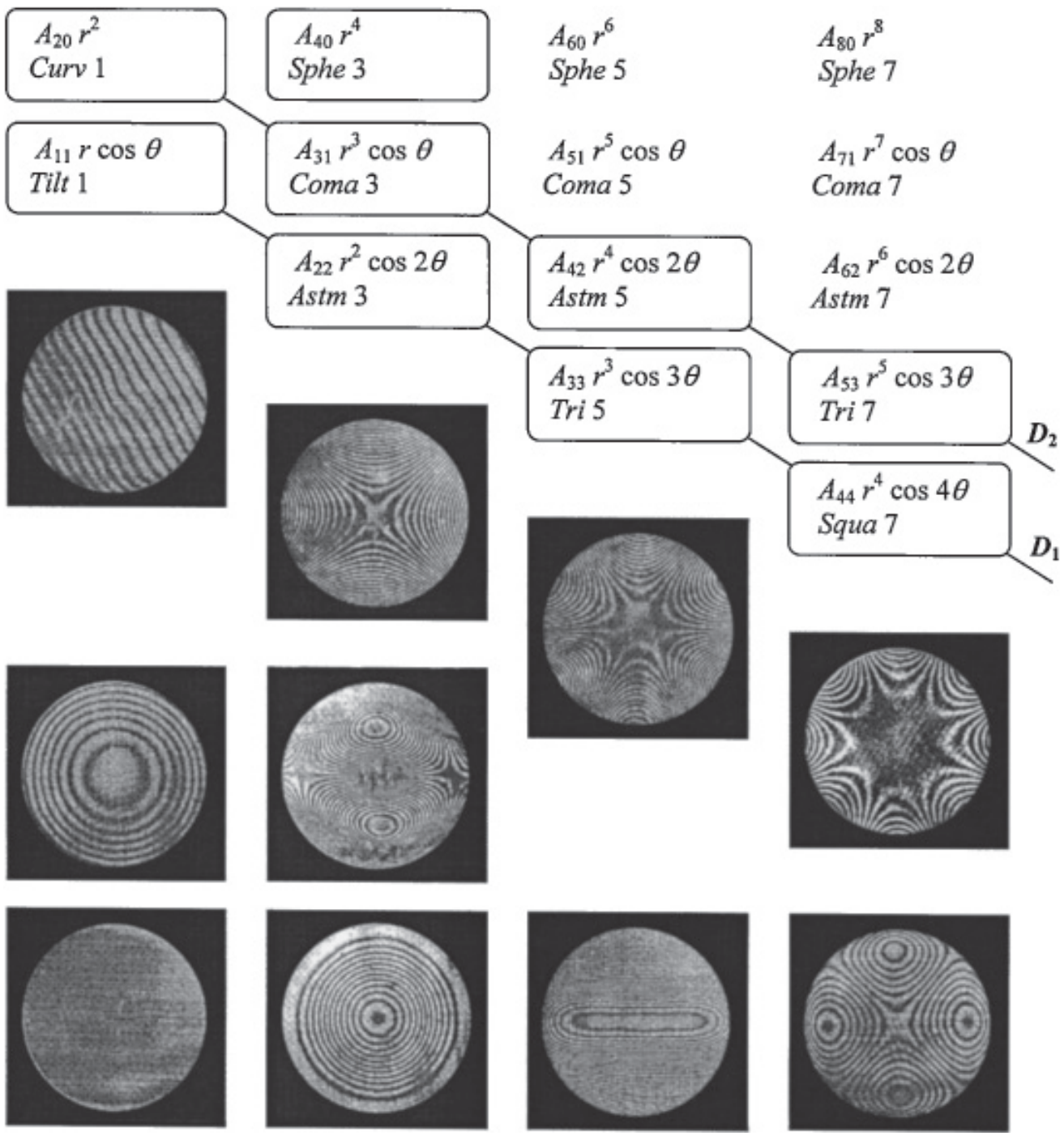

Fig. 40. Triangle matrix of Seidel optical modes. The Clebsch-Seidel modes are shown in boxes (LOOM).

### 8.2 Degenerated configurations: Monomode mirrors

Generating a single mode with an MDM equipped of $K$ radial arms generally requires use of $2K$ forces. One demonstrates from the thin plate theory that some configurations, called *degenerated*, require a maximum of $K$ forces only. A family of them is for $m = n$. Experimental results provided confirmations for $m = n = 2$ and $m = n = 3$, i.e. *Astm*3 and *Tri*5 modes.[1,16] For instance a vase form, made of quenched stainless steel FeCr13, provided pure *Astm*3 mode with four folded arms and four forces only (Fig. 42).

### 8.3 Meniscus MDMs: Keck Telescope segments

MDMs originated from a theoretical study by Lubliner and Nelson[17] for the segmented 10 m, f /1.7, Keck Telescope, the largest optical-infrared telescope. Nelson et al.[18] applied the method to the construction of 36 segments of the primary mirror. The stress figuring asphierization process of 1.8m segments used circular Zerodur meniscus. $K = 24$ radial arms—acting on the edge through stuck plates—distributed the bending moments and net shearing forces to generate the superposition of *Cv*1, *Coma*3, and *Astm*3 modes (Fig. 43).

### 8.4 Meniscus in-situ reshaping and telescope alignment control

With the construction of the 3.5m ESO-NTT, Wilson et al.[19,20] pioneered new concepts to achieve the best image quality during observations with large telescopes. The thickness of the primary mirror becomes thinner, t/D = 1/15, and thus allowed in-situ reshaping by a set of support actuators. The secondary mirror is actively set up to optimal position by avoiding decenter and tip-tilt. Wavefront sensors tracking natural stars provide the useful information for analysis. *Closed loop systems* provide efficient control drives whatever the position on the sky. The concept was then applied successfully to the four ESO 8.2m units, t/D = 1/47, of the ESO-VLT (Fig. 44).

### 8.5 Meniscus MDM: Giant reflective Schmidt LAMOST

Giant reflective Schmidt LAMOST is a quasi-meridian telescope fully dedicated to *multi-object spectroscopy*. A 4.2m clear aperture and 5° field of view confer this telescope an unprecedented *optical etendue*.[1] 4000 optics fibers are remotely positioned over a focal surface of 1.75m in diameter. The fiber outputs feed 16 double spectrographs.

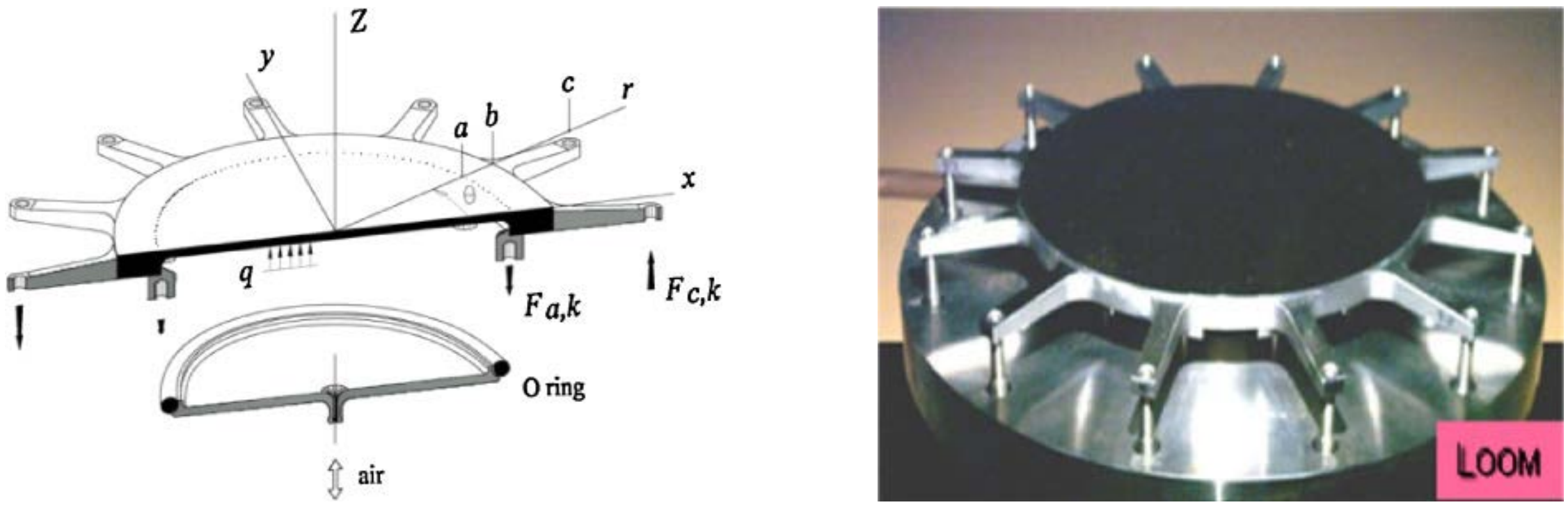

Fig. 41. Elasticity design and view of a 12-arm *vase form* MDM (LOOM).

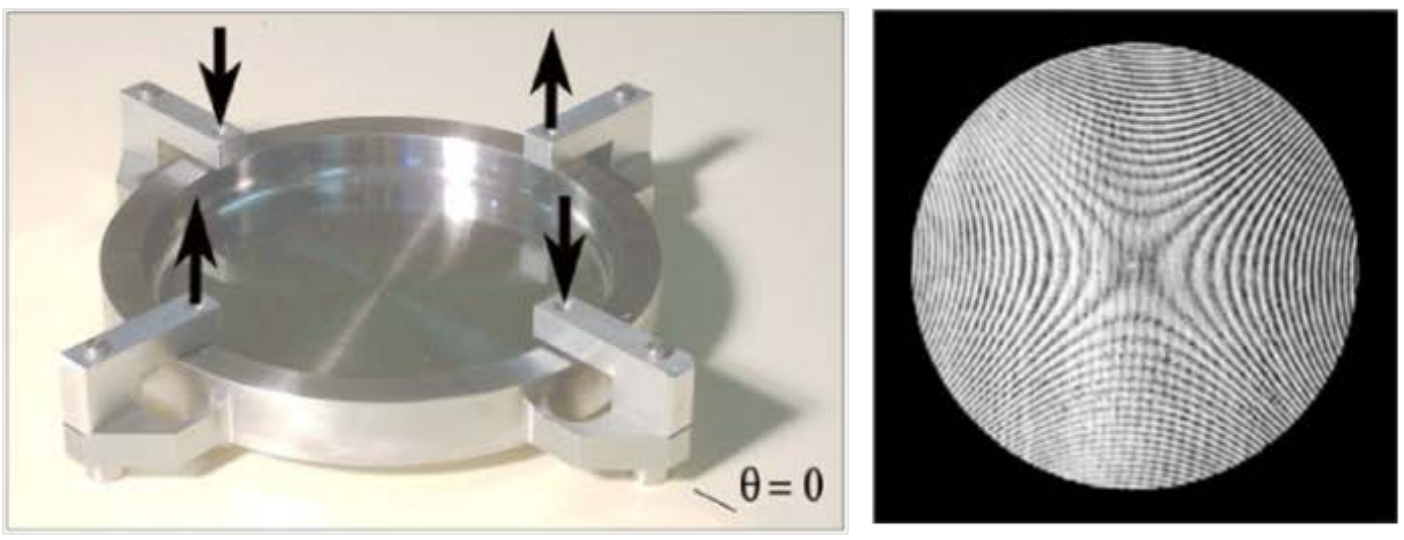

Fig. 42. Degenerated vase form configuration for $m = n = 2$ generating *Astm*3 with four forces only - instead of eight (LOOM).

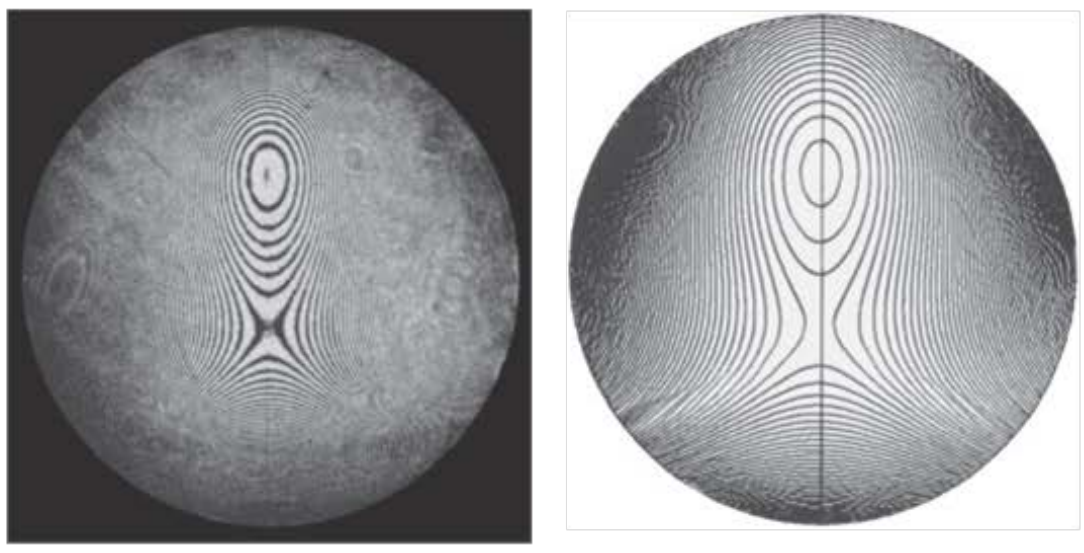

Fig. 43. Obtained and theoretical shapes of an outermost segment of the 10m Keck Telescopes (Univ. of California).

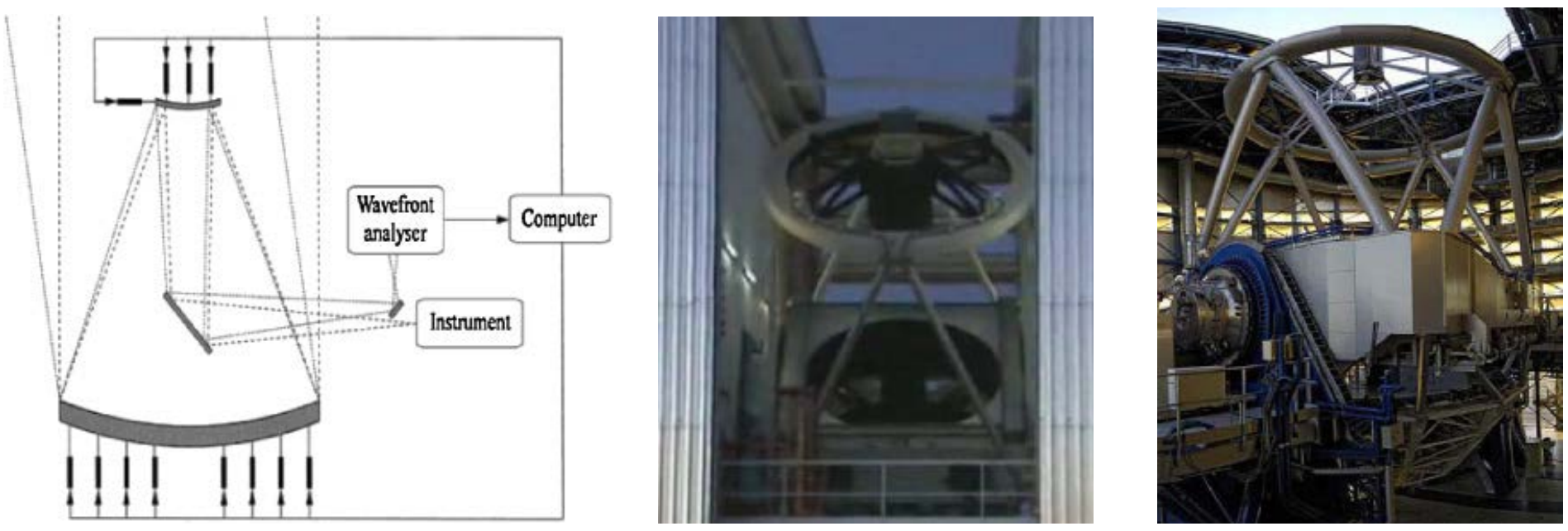

Fig. 44. Principle of telescope close-loop control—Views of NTT and VLT (ESO).

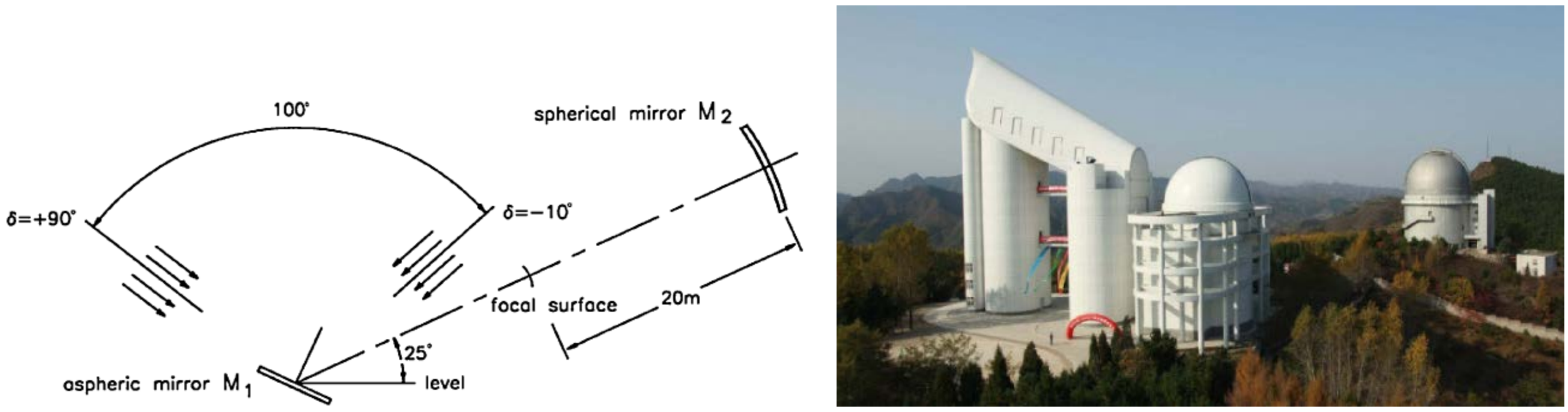

Fig. 45. Optical design of giant reflective Schmidt LAMOST. View at Xinglong Station (NAOC/CAS).

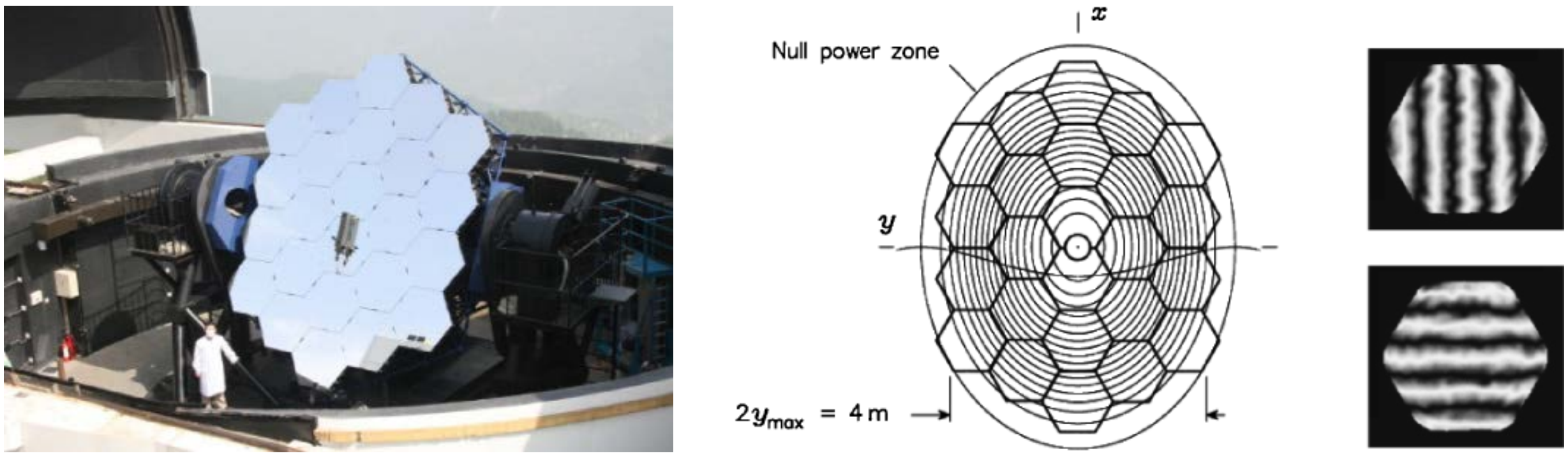

Fig. 46 LAMOST primary mirror. The ellipticity of iso-level lines of the primary mirror to aspherize is a function of the incidence angle from the sky (NIAOT & NAOC/CAS).

The concept by Wang, Su et al.[21] will provide the highest spectrum acquiring rate (Fig. 45). Active optics in LAMOST is a most extensive development ever carried out since the 24 flat segments of the primary mirror are *in-situ aspherized* as a function of the deviation angle and transit angle from the meridian (±15° for 2-h exposure time). The elliptic contour of mirror M1 (Fig. 46) corresponds to the maximum deviation from the sky. The shape and position of the M1 segments are generated by 37 force-actuators and 3 displacement-actuators.

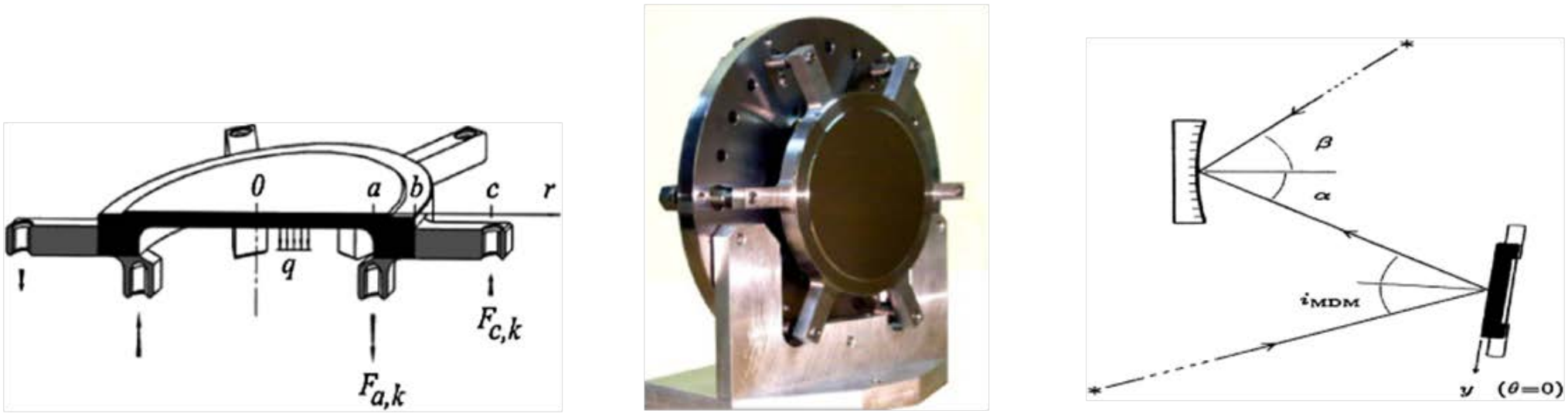

Fig. 47. Elasticity design and view of a 6-arm vase form MDM. Optical mounting for the holographic recording of aberration corrected gratings (LOOM).

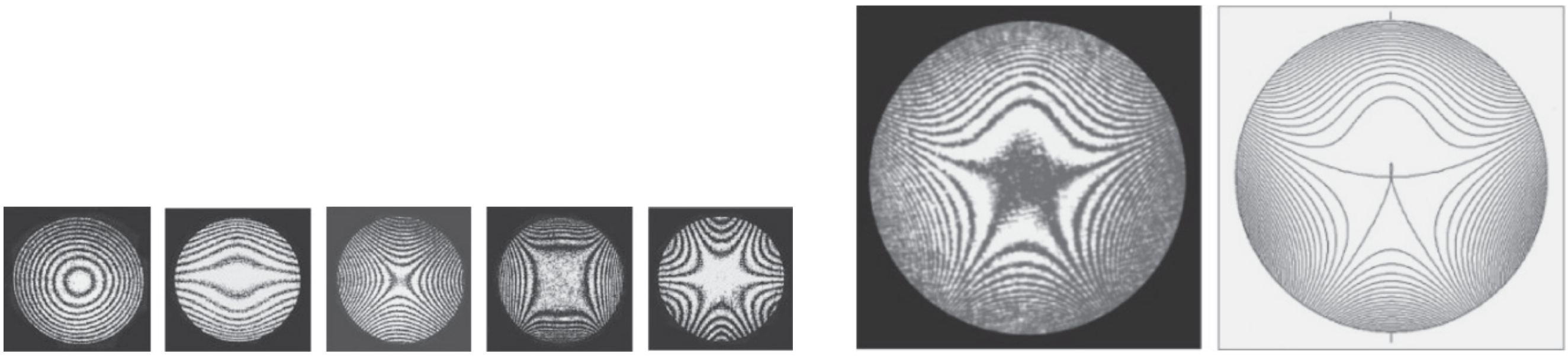

Fig. 48. (*Left*) *Cv*1, *Coma*3, *Astm*3, *Squa*5 and *Tri*5 modes generated by a 6-arm MDM. (*Right*) Mode superposition: Experimental and theoretical shape for the COS/HST holographic grating recording (LOOM).

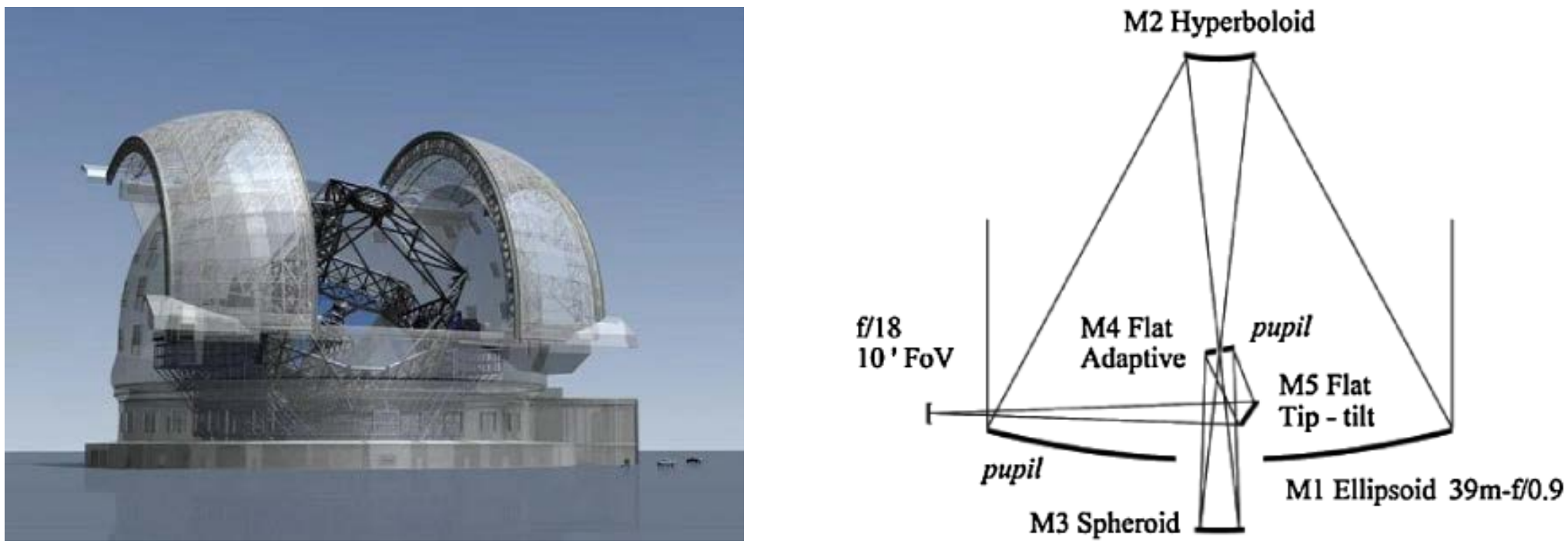

Fig. 49. The E-ELT Project. Five-mirror design including pupil transfer on adaptive mirror M4 (ESO).

displacement-actuators. LAMOST active optics systems have been conceived and developed by Su et al.[22,23] and Cui et al.[24,25]

*8.6 Aberration universal compensator for holographic grating recording—Vase form MDMs*

Universal MDM compensators for low- and high-order aberrations were proposed and developed for holographic recording of concave gratings. Adopting a vase form, this was applied to the Cosmic Origins Spectrograph – COS – of the Hubble Space Telescope. Depict of nominal results (Figs. 47 and 48) our proposal did not finalized for COS. It would have led to a gain of 1.5 in magnitude and four in spectral resolution. The recording compensator was a 6-arm vase form MDM in quenched alloy.[1,26,27]

*8.7 Aspherization process for the E-ELT segments*

We presently develop a new industrialization procedure based on the *vase form* MDM concept[1] for aspherizing outermost M1 segments of the European Extremely Large Telescope (E-ELT) project by European Southern Observatory. This 40-m design is a five-mirror train which *includes a pupil transfer* for ground layer corrections on adaptive mirror M4 and field stabilization mirror M5 (Fig. 49).

Mirror M1 will be made of ~900 hexagonal segments, 1.35-m in diagonal, which form an f/0.9 slightly elongated ellipsoid. The aspherization process is developed by Ferrari et al.[28], Laslandes et al.[29], Hugot and Floriot. It will use a 12-arm MDM stressing harness[1] where a Zerodur segment is linked to an active ring during aspherization (Fig. 50).

## 9. Active Optics and X-ray Optics

Due to the extremely short wavelengths of X-rays the fabrication of X-ray mirrors requires *super-smooth reflective surfaces*. This is without doubt one of the most important features because slope errors due to ripples entail absorption and scattering effects which may severely degrade the performances. One usually admits a surface roughness not exceeding 2-3Â. For this reason active optics figuring with rigid laps and in-situ active optics would greatly improve the performances of tubular mirrors although not much used up to now except for long stripe-mirrors.

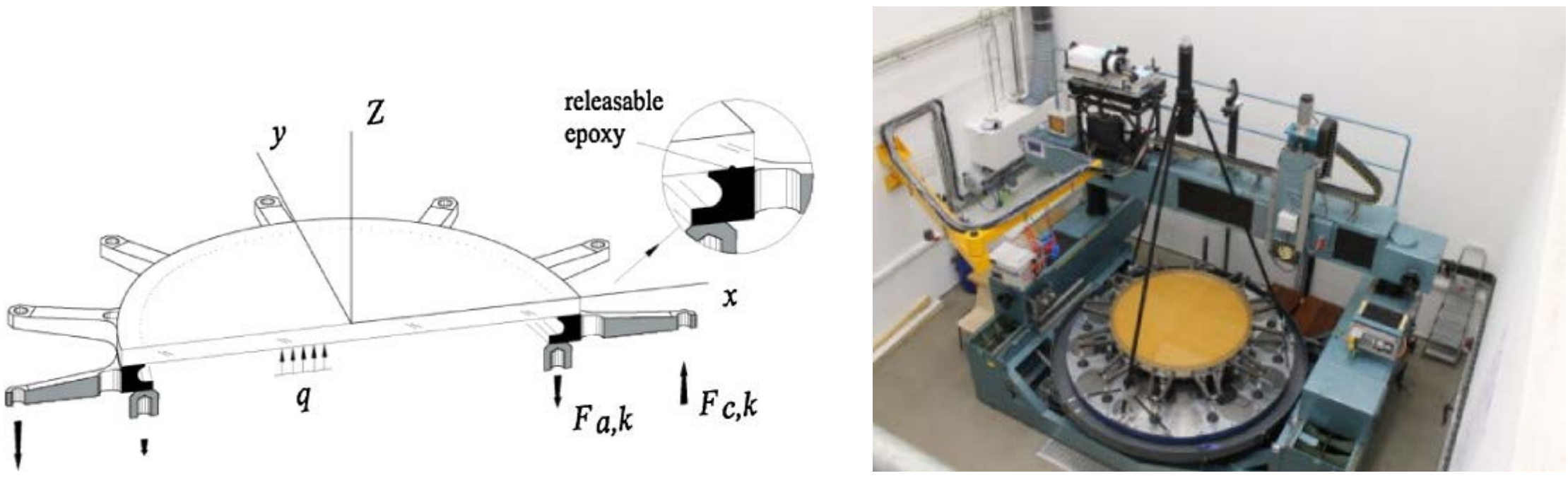

Fig. 50. MDM assembly for the asphierization process of E-ELT segments. View of the platform at LOOM.

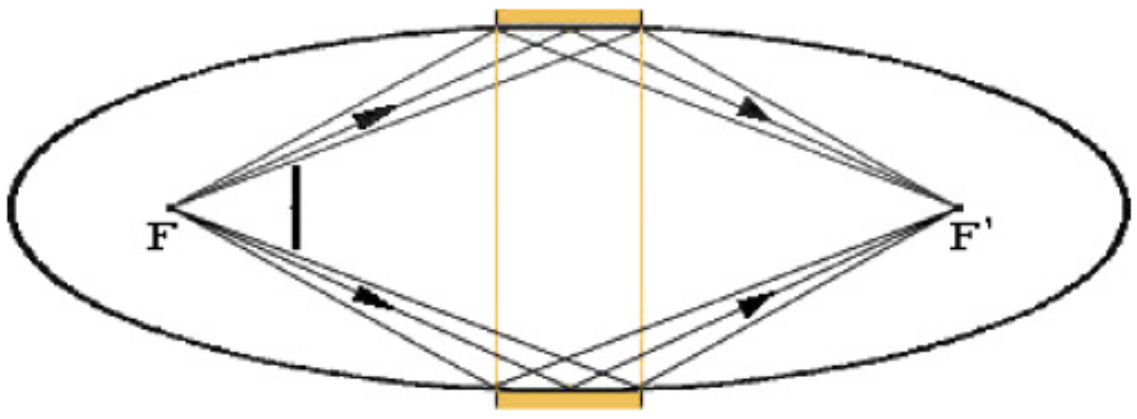

Fig. 51. Conjugate focii of an elongated ellipsoid.

*9.1 Grazing incidence: X-ray relay mirrors*

For instance the elasticity theory of tubular shells would allow obtaining a tubular section of a stigmatic elongated ellipsoid. The shape is required to build accurate relay mirrors (Fig. 51).

If one considers a *monotonic sign extension or compression* of a tubular shell, caused by a uniform load *q* applied at the surface of the shell, one shows that the radial thickness distribution $T\,(\chi)$ can be selected to remain of finite value. This avoids distributions which exhibit local infinite thicknesses. For generating a given polynomial flexure along the shell, i.e., along -axis, this theory provides two solution families (Lemaitre[1,30]) :

⟹ *If the first family is represented by a load* $q > 0$ *and thickness* $T(\chi)\ \uparrow$ *from* $\chi = 0$ *to outer ends* $\pm\chi_{max}$ *, then a second family exists with* $q < 0$ *and* $T\,(\chi) \downarrow$ *from* $\chi = 0$ *to* $\pm\chi_{max}$ .

As a result of this law, one displays thickness distribution two families which generate a parabolic flexure (curvature mode) and a 4th-degree flexure by uniform loading *q* positive and negative (Fig. 52). A combined distribution of these thicknesses easily provide the convergence mode and stigmatism mode for the elasticity design of elongated ellipsoids usable as X-ray relay mirrors.[1)]

*9.2 X-ray tubular telescope mirrors and sine condition*

Future high angular resolution X-ray two-mirror tubular telescopes will evolve from the classical paraboloid-hyperboloid (PH) optical design (Fig. 53) towards a Wolter–Scwhartzchild (WS) design—as derived by Chase VanSpeybroeck[31)]—which fully satisfies Abbe's sine condition (Fig. 54). Due the extremely short wavelength and the foreseen new goal of 0.1 arcsec in resolution it is obvious that the implementation of active optics will be necessary.

One has developed an elasticity theory of weakly conical shells[1,30)] and applied it to the optical design of such high-resolution telescope mirrors. Analytic results show that radial flexure distributions can be generated along the shell with a monotonic sign. This latter feature avoids the difficulty of a distribution with infinite thicknesses.

Let us represent the distribution of the radial flexural extension—or compression—W( $\chi$ ) along $\chi$-axis of the tube as the series

$$W = \pm\,\alpha^2 + \Sigma^N{}_{n=1,2,3\ldots}\ A_n\,\chi^n$$

where $\chi$ is the axis of the mirror which ends at $[-\beta;\,\beta]$ , $\alpha$ a constant to be set up for obtaining a monotonic sign of the flexure all along the mirror, and $A_n$ coefficients determined by the mirror shape deviation to a best fit cone or circle. The result of weakly conical shell theory provides the thickness $T\,(\chi)$ as easily derived from the flexure W($\chi$) by a linear product law (GL[1,30)]):

$$T\,(\chi)\,W(\chi) = C(1\text{-}2i\chi)/(1\text{-}t_0/a_0), \quad \chi\in[-\beta,\beta]$$

where $C$ is a constant (from load $q$, Young modulus $E$), $i$ is the mirror slope at $\chi = 0$, $t_0$ is the shell thickness at $\chi = 0$, $a_0$ is the shell radius respectively at $\chi$=0.

This law has been applied to the determination of thickness distribution $T\,(\chi)$ for mirror #1 of the WS design in Fig. 55. The result is that the thickness profile $T\,(\chi)$ also shows two inflexion points very similarly to that of the optical design (Fig. 55).

*9.3 Segmented X-ray telescope mirrors*

Similarly as large ground-based telescopes in the visible and infrared, future large space-based X-ray telescopes will require use of segmented and active mirrors. Current projects under investigation, i.e., Constellation X-ray Mission (NASA) and Xeus (ESA), mostly have architectures based on formation flyers.

From above results of a monolithic tubular shell, active optics can be applied to the asphierization of an X-ray segment of tubular telescope. The boundaries of this segment are derived from the stress distributions of a truncated part of a full shell (Fig. 56).[1] Three boundary conditions must be satisfied for the aspherization of a segment through a harness.

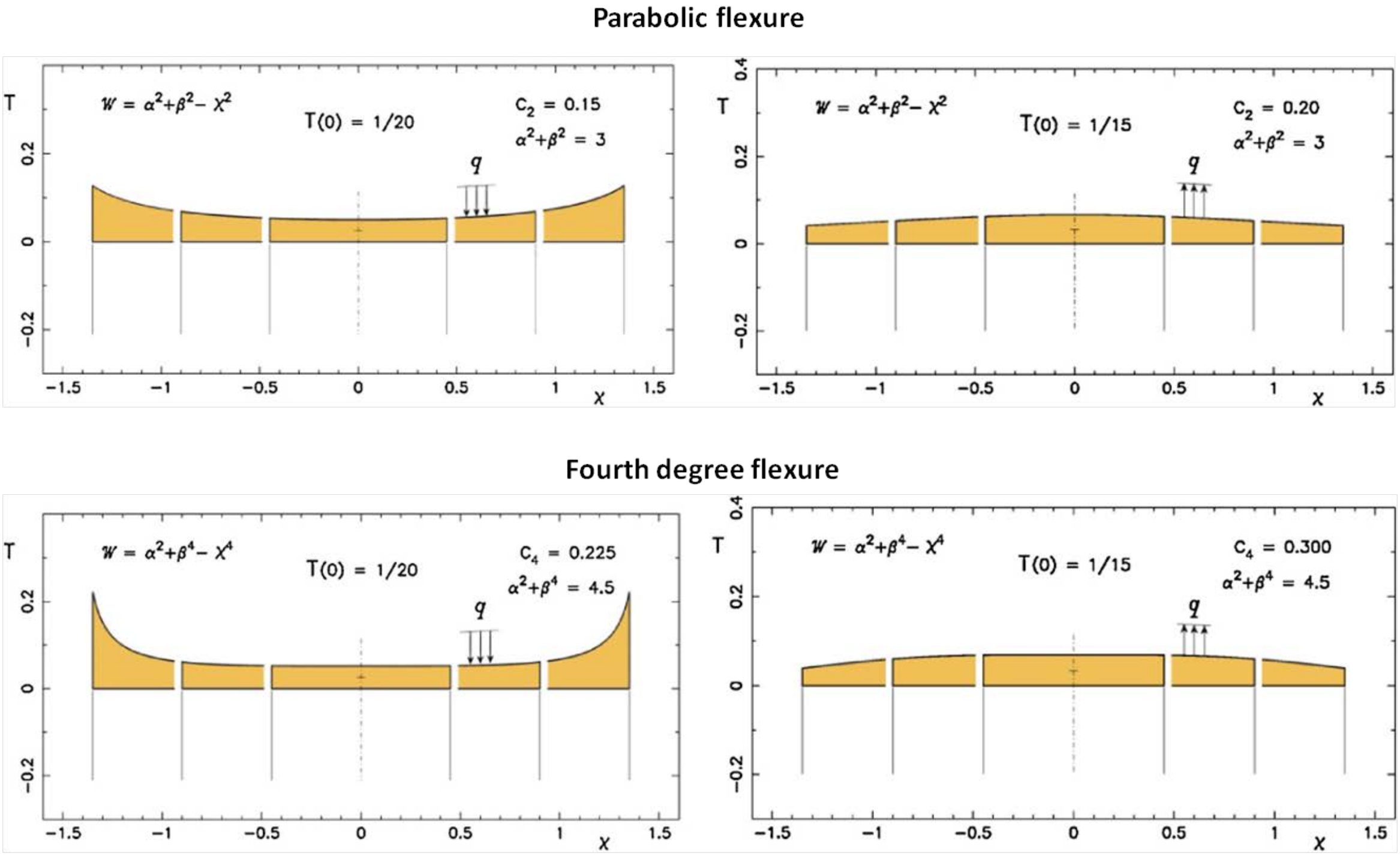

Fig. 52. Thickness distributions for a 2nd- and 4th-degree flexure along three cylindrical shells submitted to uniform load *q*. Opposite signs of *q*—left and right figures—show the two families that generate the same flexure.The thickness variations are greatly exaggerated for clarity.

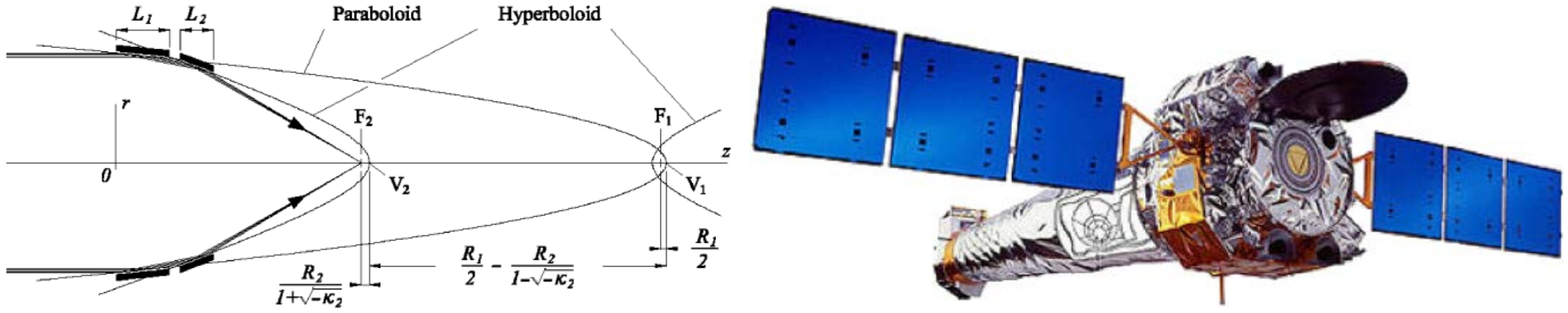

Fig. 53. (*Left*)Basic paraboloid-hyperboloid (PH) design of X-ray two-mirror telescope. Chandra X-ray. (*Right*) Observatory launched in 1999 is a PH design (NASA).

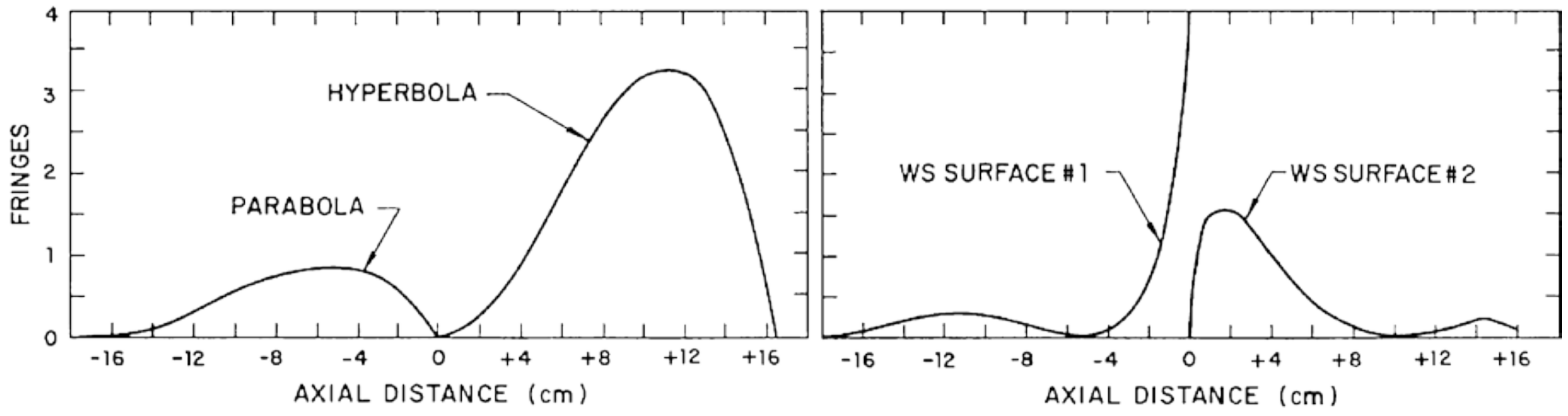

Fig. 54. Comparison of X-ray two-mirror telescope designs: PH design versus WS design. Mirror shapes with respect to best fit circle. Wolter–Schwarzschild (WS) form satisfies the sine condition (Chase & Van Speybroeck).

This harness must fulfill the following properties:

C1: facets at $\pm\psi/2$ supported by normal pressure $p$,
C2: facets free to slide in the dihedral planes $\pm\psi/2$,
C3: receive reaction $R_p$ to load $q$ at largest facet end.

The accurate achievement of these conditions does not present major technical difficulties.

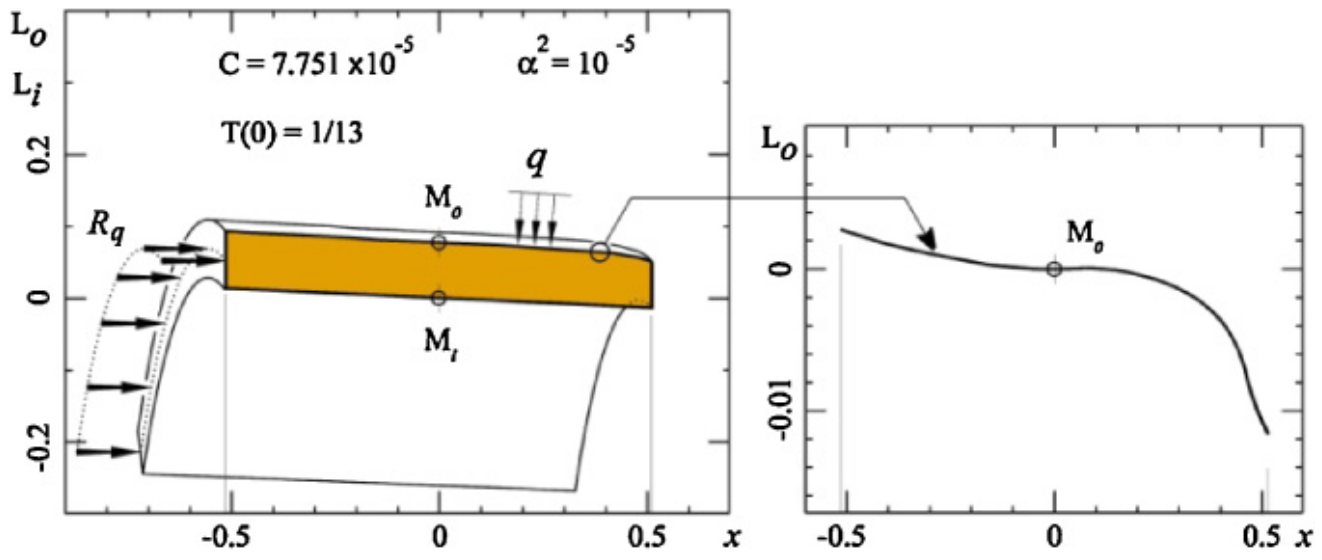

Fig. 55. Weakly conical shell and thickness distribution $T(\chi)$ showing two inflexion points as for the optical shape of mirror #1 of WS primary mirror in Fig. 54. Aspherization by uniform load $q$ from best fit circle.

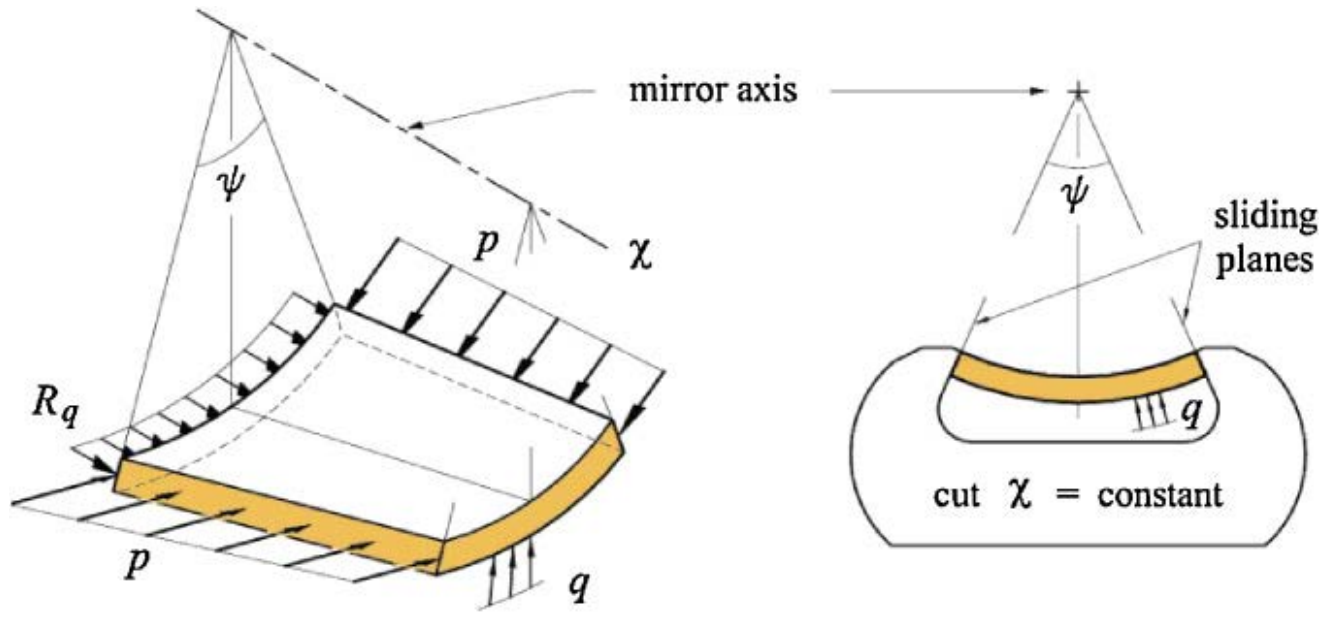

Fig. 56. X-ray telescope segment: Configuration for aspherization by stress figuring and uniform loading. Basic harness (*right*) satisfying the boundary conditions C1, C2 and C3.

## 10. Conclusions

Since pioneer works, almost 50 years ago, active optics methods provided fascinating imaging quality in astronomy either in high angular resolution or limiting magnitude detection.

These methods and developments are key features to bring astronomers new powerful telescopes and associated astronomical instrumentations.